\documentclass[12pt]{article}
\usepackage{a4wide,epsfig,epsf}

\newcommand{\re}{\mathop{\mathrm{Re}}\nolimits}
\newcommand{\arcosh}{\mathop{\mathrm{arcosh}}\nolimits}

\newcommand{\li}{\mathop{\mathrm{Li}_2}\nolimits}
\newcommand{\agt}{\,\rlap{\lower 3.5 pt \hbox{$\mathchar \sim$}} \raise 1pt
 \hbox {$>$}\,}
\newcommand{\alt}{\,\rlap{\lower 3.5 pt \hbox{$\mathchar \sim$}} \raise 1pt
 \hbox {$<$}\,}

\catcode`@=11
\newcount\@tempcntc
\def\@citex[#1]#2{\if@filesw\immediate\write\@auxout{\string\citation{#2}}\fi
  \@tempcnta\z@\@tempcntb\m@ne\def\@citea{}\@cite{\@for\@citeb:=#2\do
    {\@ifundefined
       {b@\@citeb}{\@citeo\@tempcntb\m@ne\@citea\def\@citea{,}{\bf ?}\@warning
       {Citation `\@citeb' on page \thepage \space undefined}}%
    {\setbox\z@\hbox{\global\@tempcntc0\csname b@\@citeb\endcsname\relax}%
     \ifnum\@tempcntc=\z@ \@citeo\@tempcntb\m@ne
       \@citea\def\@citea{,}\hbox{\csname b@\@citeb\endcsname}%
     \else
      \advance\@tempcntb\@ne
      \ifnum\@tempcntb=\@tempcntc
      \else\advance\@tempcntb\m@ne\@citeo
      \@tempcnta\@tempcntc\@tempcntb\@tempcntc\fi\fi}}\@citeo}{#1}}
\def\@citeo{\ifnum\@tempcnta>\@tempcntb\else\@citea\def\@citea{,}%
  \ifnum\@tempcnta=\@tempcntb\the\@tempcnta\else
   {\advance\@tempcnta\@ne\ifnum\@tempcnta=\@tempcntb \else \def\@citea{--}\fi
    \advance\@tempcnta\m@ne\the\@tempcnta\@citea\the\@tempcntb}\fi\fi}
\catcode`@=12

\begin{document}

\title{
\vskip-3cm{\baselineskip14pt
\centerline{\normalsize DESY 04-080\hfill ISSN 0418-9833}
\centerline{\normalsize LPSC 04-017\hfill}
\centerline{\normalsize hep-ph/0407014\hfill}
\centerline{\normalsize June 2004\hfill}
}
\vskip1.5cm
$J/\psi$ plus jet associated production in two-photon collisions at
next-to-leading order}
\author{
{\sc M. Klasen,${}^a$ B.A. Kniehl,${}^b$ L.N. Mihaila,${}^b$
M. Steinhauser${}^b$}\\
{\normalsize ${}^a$ Universit\'e Grenoble I,
Laboratoire de Physique Subatomique et de Cosmologie,}\\
{\normalsize 53 Avenue des Martyrs, 38026 Grenoble, France}\\
{\normalsize ${}^b$ II. Institut f\"ur Theoretische Physik,
Universit\"at Hamburg,}\\
{\normalsize Luruper Chaussee 149, 22761 Hamburg, Germany}}

\date{}

\maketitle

\thispagestyle{empty}

\begin{abstract}
We calculate the cross section of $J/\psi$ plus jet inclusive production in
$\gamma\gamma$ collisions at next-to-leading order within the factorization
formalism of nonrelativistic quantum chromodynamics (NRQCD) focusing on direct
photoproduction.
Apart from direct $J/\psi$ production, we also include the feed-down from
directly-produced $\chi_{cJ}$ and $\psi^\prime$ mesons.
We discuss the analytical calculation, in particular the treatment of the
various types of singularities and the NRQCD operator renormalization, in some
detail.
We present theoretical predictions for the future $e^+e^-$ linear collider
TESLA, taking into account both brems- and beamstrahlung.

\medskip

\end{abstract}

\newpage

\section{Introduction}
\label{sec:one}

Since the discovery of the $J/\psi$ meson in 1974, charmonium has provided a
useful laboratory for quantitative tests of quantum chromodynamics (QCD) and,
in particular, of the interplay of perturbative and nonperturbative phenomena.
The factorization formalism of nonrelativistic QCD (NRQCD) \cite{cas,bbl}
provides a rigorous theoretical framework for the description of
heavy-quarkonium production and decay.
This formalism implies a separation of short-distance coefficients, which can 
be calculated perturbatively as expansions in the strong-coupling constant
$\alpha_s$, from long-distance matrix elements (MEs), which must be extracted
from experiment.
The relative importance of the latter can be estimated by means of velocity
scaling rules; i.e., the MEs are predicted to scale with a definite power of
the heavy-quark ($Q$) velocity $v$ in the limit $v\ll1$.
In this way, the theoretical predictions are organized as double expansions in
$\alpha_s$ and $v$.
A crucial feature of this formalism is that it takes into account the complete
structure of the $Q\overline{Q}$ Fock space, which is spanned by the states
$n={}^{2S+1}L_J^{(a)}$ with definite spin $S$, orbital angular momentum
$L$, total angular momentum $J$, and colour multiplicity $a=1,8$.
In particular, this formalism predicts the existence of colour-octet (CO)
processes in nature.
This means that $Q\overline{Q}$ pairs are produced at short distances in
CO states and subsequently evolve into physical, colour-singlet (CS) quarkonia
by the nonperturbative emission of soft gluons.
In the limit $v\to0$, the traditional CS model (CSM) \cite{ber} is recovered.
The greatest triumph of this formalism was that it was able to correctly 
describe \cite{bra} the cross section of inclusive charmonium
hadroproduction measured in $p\overline{p}$ collisions at the Fermilab
Tevatron \cite{abe}, which had turned out to be more than one order of
magnitude in excess of the theoretical prediction based on the CSM.

Apart from this phenomenological drawback, the CSM also suffers from severe
conceptual problems indicating that it is incomplete.
These include the presence of logarithmic infrared singularities in the
${\cal O}(\alpha_s)$ corrections to $P$-wave decays to light hadrons and in
the relativistic corrections to $S$-wave annihilation \cite{bar}, and the lack
of a general argument for its validity in higher orders of perturbation
theory.
While the $k_T$-factorization \cite{sri} and hard-comover-scattering
\cite{hoy} approaches manage to bring the CSM prediction much closer to the
Tevatron data, they do not cure the conceptual defects of the CSM.
The colour evaporation model \cite{cem}, which is intuitive and useful for
qualitative studies, also significantly improves the description of the
Tevatron data as compared to the CSM \cite{sch}.
However, it does not account for the process-specific weights of the CS and CO
contributions, but rather assumes a fixed ratio of $1:7$.
In this sense, a coequal alternative to the NRQCD factorization formalism is 
presently not available.

In order to convincingly establish the phenomenological significance of the
CO processes, it is indispensable to identify them in other kinds of
high-energy experiments as well.
Studies of charmonium production in $ep$ photoproduction \cite{cac}, $ep$
\cite{fle,ep} and $\nu N$ \cite{pet,nun} deep-inelastic scattering (DIS),
$e^+e^-$ annihilation in the continuum \cite{che}, $Z$-boson decays
\cite{keu}, $\gamma\gamma$ collisions \cite{ma,npb,god,gg}, and $b$-hadron
decays \cite{ko} may be found in the literature; for reviews, see
Ref.~\cite{yua}.
Furthermore, the polarization of $\psi^\prime$ mesons produced directly
\cite{ben} and of $J/\psi$ mesons produced promptly \cite{bkl,kni}, i.e.,
either directly or via the feed-down from heavier charmonia, which also
provides a sensitive probe of CO processes, was investigated.
Until recently, none of these studies was able to prove or disprove the NRQCD
factorization hypothesis.
However, H1 data of $ep\to e+J/\psi+X$ in DIS at the DESY Hadron Electron Ring
Accelerator (HERA) \cite{h1} and DELPHI data of $\gamma\gamma\to J/\psi+X$ at
the CERN Large Electron Positron Collider (LEP2) \cite{delphi} provide first
independent evidence for it \cite{ep,gg}.

The verification of the NRQCD factorization hypothesis is presently hampered
both from the theoretical and experimental sides.
On the one hand, the theoretical predictions to be compared with existing
experimental data are, apart from very few exceptions \cite{kra,man,gre}, of
lowest order (LO) and thus suffer from considerable uncertainties, mostly from
the dependences on the renormalization and factorization scales and from the
lack of information on the nonperturbative MEs.
On the other hand, the experimental errors are still rather sizeable.
The latter will be dramatically reduced with the upgrades of HERA (HERA II)
and the Tevatron (Run II) and with the advent of CERN LHC and hopefully a
future $e^+e^-$ linear collider (LC) such as the TeV-Energy Superconducting
Linear Accelerator (TESLA), which is presently being designed and planned at
DESY.
On the theoretical side, it is necessary to calculate the
next-to-leading-order (NLO) corrections to the hard-scattering cross sections
and to include the effective operators which are suppressed by higher powers
in $v$.

In this paper, we take a first step in this comprehensive research programme,
by studying the inclusive production of $J/\psi$ mesons in high-energy
$\gamma\gamma$ collisions.
As mentioned above, this process was studied at LEP2 \cite{delphi}, where the
photons originated from hard initial-state bremsstrahlung.
At high-energy $e^+e^-$ LCs, an additional source of hard photons is provided
by beamstrahlung, the synchrotron radiation emitted by one of the colliding
bunches in the field of the opposite bunch.
The highest possible photon energies with large enough luminosity may be
achieved by converting the $e^+e^-$ LC into a $\gamma\gamma$ collider via
back-scattering of high-energetic laser light off the electron and positron
beams.

In order for a $J/\psi$ meson to acquire finite transverse momentum ($p_T$),
it must be produced together with another particle or a hadron jet ($j$).
From coupling and phase-space considerations it is evident that $J/\psi$ plus
jet associated production yields the dominant contribution.
In the following, we thus consider the process $\gamma\gamma\to J/\psi+j+X$,
where $X$ denotes the hadronic remnant possibly including a second jet.
Here, we take $j$ and $X$ to be free of charm assuming that charmed hadrons or
charmonia besides the $J/\psi$ meson would be detectable.
The process $\gamma\gamma\to J/\psi+\gamma+X$, where $\gamma$ represents a
prompt photon, will be considered in a forthcoming publication \cite{kkms}.

The incoming photons can interact either directly with the quarks
participating in the hard-scattering process (direct photoproduction) or via
their quark and gluon content (resolved photoproduction).
Thus, the process $\gamma\gamma\to J/\psi+j+X$ receives contributions from the
direct, single-resolved, and double-resolved channels.
All three contributions are formally of the same order in the perturbative
expansion.
This may be understood by observing that the parton density functions (PDFs)
of the photon have a leading behaviour proportional to
$\alpha\ln(M^2/\Lambda_{\rm QCD}^2)\propto\alpha/\alpha_s$, where $\alpha$ is
the fine-structure constant, $M$ is the factorization scale, and
$\Lambda_{\rm QCD}$ is the asymptotic scale parameter of QCD.
In the following, we focus our attention on the direct channel.
The other channels are left for future work.

The $J/\psi$ mesons can be produced directly; or via radiative or hadronic
decays of heavier charmonia, such as $\chi_{cJ}$ and $\psi^\prime$ mesons; or
via weak decays of $b$ hadrons.
The respective decay branching fractions are
$B(\chi_{c0}\to J/\psi+\gamma)=(1.02\pm0.17)\%$,
$B(\chi_{c1}\to J/\psi+\gamma)=(31.6\pm3.2)\%$,
$B(\chi_{c2}\to J/\psi+\gamma)=(18.7\pm2.0)\%$,
$B(\psi^\prime\to J/\psi+X)=(55.7\pm2.6)\%$, and
$B(B\to J/\psi+X)=(1.15\pm0.06)\%$ \cite{pdg}.
The $b$ hadrons can be detected by looking for displaced decay vertices with
dedicated vertex detectors, and the $J/\psi$ mesons originating from their
decays can thus be treated separately.
Therefore and because of the smallness of $B(B\to J/\psi+X)$, $J/\psi$
production through $b$-hadron decay is not considered here.
The cross sections of the four residual indirect production channels may be
approximated by multiplying the direct-production cross sections of the
respective intermediate charmonia with their decay branching fractions to
$J/\psi$ mesons.

To summarize, the goal of the present analysis is to calculate the inclusive
cross section of $\gamma\gamma\to J/\psi+X$ in direct photoproduction for
finite values of $p_T$ at NLO within NRQCD allowing for the $J/\psi$ meson to
be promptly produced.
The LO result, also including the single- and double-resolved contributions,
may be found in Refs.~\cite{god,gg} and the references cited therein.
The leading relativistic correction, which originates from the $c\overline{c}$
Fock state $n={}^1\!P_1^{(8)}$ and is of ${\cal O}(v^4)$ relative to the LO
result, was evaluated in Ref.~\cite{npb}.
In that paper, also $J/\psi$ plus dijet associated production in two-photon
collisions was studied for direct photoproduction and direct $J/\psi$
production.
The $2\to3$ partonic cross sections obtained therein constitute a starting
point for the evaluation of the real radiative corrections in our present
study.
They have to be complemented with the corresponding results for direct
$\chi_{cJ}$ production.
Furthermore, the kinematic cuts imposed in Ref.~\cite{npb} to separate the
final-state objects have to be removed.
This yields soft and collinear singularities, which are collectively denoted 
as infrared (IR) singularities.
On top of this, the virtual radiative corrections, which arise from $2\to2$
Feynman diagrams\footnote{%
Here, the $c\overline{c}$ bound state is considered as one particle.}
involving closed loops, have to be added.
They involve IR, ultraviolet (UV), and Coulomb singularities.
The cancellation of all these singularities is nontrivial and requires the
UV renormalization of masses, couplings, and wave-functions; the redefinition
of NRQCD MEs so as to absorb IR and Coulomb singularities; the factorization
of initial-state collinear singularities in photon PDFs; and the operation of
the Kinoshita-Lee-Nauenberg theorem \cite{kln} on the cancellation of 
final-state IR singularities.
Apart from being of general phenomenological relevance, our analysis should
thus also be of conceptual interest for the theoretical heavy-quarkonium
community.
After all, this is the first time that the full NLO corrections are evaluated
for an inclusive $2\to2$ process within the NRQCD framework.

This paper is organized as follows.
In Section~\ref{sec:ana}, we describe our analytical calculation in some
detail.
Specifically, we discuss the structure of the various types of singularities
and the mechanisms by which they are removed.
Lengthy expressions are relegated to the Appendix.
In Section~\ref{sec:num}, we present our numerical results appropriate for
 the $e^+e^-$ mode of TESLA, and discuss their phenomenological
implications.
Our conclusions are summarized in Section~\ref{sec:con}.

\section{Analytic results}
\label{sec:ana}

We start this section with a few general remarks.
In our analytic calculation, we take the colour gauge group to be SU($N_c$)
with a generic value of $N_c$, which is put equal to 3 in our numerical 
analysis. 
Colour factors appearing in our formulas include
$T_F=1/2$, $C_F=\left(N_c^2-1\right)/(2N_c)$, $C_A=N_c$, and 
$B_F=\left(N_c^2-4\right)/(4N_c)$.
We work in the fixed-flavour-number scheme, with $n_f=3$ active quark flavours
$q=u,d,s$, which we treat as massless.
The charm quark $c$ and antiquark $\overline{c}$, with mass $m$, only appear
in the final state.
We denote the fractional electric charge of quark $q$ by $e_q$.

\begin{table}[t]
\begin{center}
\begin{tabular}{|c|cc|}
\hline\hline
$k$ & $J/\psi$, $\psi^\prime$ & $\chi_{cJ}$ \\
\hline
3 & ${}^3\!S_1^{(1)}$ & --- \\
5 & --- & ${}^3\!P_J^{(1)}$, ${}^3\!S_1^{(8)}$ \\
7 & ${}^1\!S_0^{(8)}$, ${}^3\!S_1^{(8)}$, ${}^3\!P_J^{(8)}$ & --- \\
\hline\hline
\end{tabular}
\caption{Values of $k$ in the velocity-scaling rule
$\left\langle{\cal O}^H[n]\right\rangle\propto v^k$ for the leading
$c\overline{c}$ Fock states $n$ pertinent to
$H=J/\psi,\chi_{cJ},\psi^\prime$.}
\label{tab:vsr}
\end{center}
\end{table}

The $c\overline{c}$ Fock states contributing at LO in $v$ are specified for
$H=J/\psi,\chi_{cJ},\psi^\prime$ in Table~\ref{tab:vsr}.
Their MEs satisfy the multiplicity relations
\begin{eqnarray}
\left\langle{\cal O}^{\psi(nS)}\left[{}^3\!P_J^{(8)}\right]\right\rangle
&=&(2J+1)
\left\langle{\cal O}^{\psi(nS)}\left[{}^3\!P_0^{(8)}\right]\right\rangle,
\nonumber\\
\left\langle{\cal O}^{\chi_{cJ}}\left[{}^3\!P_J^{(1)}\right]\right\rangle
&=&(2J+1)
\left\langle{\cal O}^{\chi_{c0}}\left[{}^3\!P_0^{(1)}\right]\right\rangle,
\nonumber\\
\left\langle{\cal O}^{\chi_{cJ}}\left[{}^3\!S_1^{(8)}\right]\right\rangle
&=&(2J+1)
\left\langle{\cal O}^{\chi_{c0}}\left[{}^3\!S_1^{(8)}\right]\right\rangle,
\label{eq:mul}
\end{eqnarray}
which follow to LO in $v$ from heavy-quark spin symmetry.

We employ dimensional regularization with $d=4-2\epsilon$ space-time
dimensions to handle the UV and IR singularities, and we introduce a 't~Hooft
mass $\mu$ and a factorization mass $M$ as unphysical scales.
We formally distinguish between UV and IR poles, which we denote as
$1/\epsilon_{\rm UV}$ and $1/\epsilon_{\rm IR}$, respectively.
We apply the projection method of Refs.~\cite{gre,kap}, which is equivalent to
the $d$-dimensional matching procedure of Ref.~\cite{bch}, in order to extract
the short-distance coefficients that multiply the MEs.
However, in order to conform with common standards, we adopt the
normalizations of the MEs from Ref.~\cite{bbl} rather than from
Refs.~\cite{man,gre}; i.e., the MEs include spin and colour average factors.

\begin{figure}[ht]
\begin{center}
\epsfig{figure=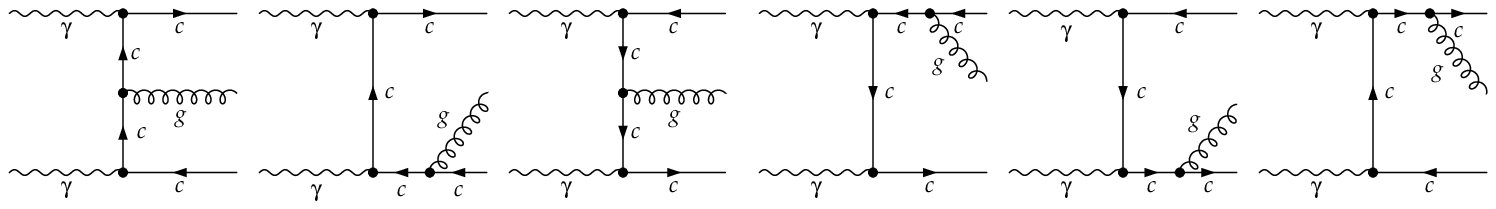,width=\textwidth,
bbllx=0pt,bblly=385pt,bburx=432pt,bbury=485pt}
\caption{Tree-level Feynman diagrams pertinent to the partonic
subprocess~(\ref{eq:ccg}).}
\label{fig:ccg}
\end{center}
\end{figure}

There is only one partonic subprocess at LO, namely
\begin{equation}
\gamma(k_1)+\gamma(k_2)\to 
c\overline{c}\left[{}^3\!S_1^{(8)}\right](p)+g(k_3),
\label{eq:ccg}
\end{equation}
which is mediated by the diagrams depicted in Fig.~\ref{fig:ccg}.
The analogous process with $n={}^1\!P_1^{(8)}$ yields a relativistic 
correction of ${\cal O}(v^4)$ and was studied in Ref.~\cite{npb}.
With the four-momentum assignments indicated within the parentheses in 
Eq.~(\ref{eq:ccg}), we have $k_1+k_2=p+k_3$, $k_1^2=k_2^2=k_3^2=0$, and
$p^2=4m^2$.
We define the Mandelstam variables as $s=(k_1+k_2)^2$, $t=(k_1-p)^2$, and
$u=(k_2-p)^2$, so that $s+t+u=4m^2$.
For convenience, we also introduce
\begin{equation}
s_1=s-4m^2,\qquad
t_1=t-4m^2,\qquad
u_1=u-4m^2.
\label{eq:stu}
\end{equation}

In the NLO analysis, we need to evaluate the cross section of process
(\ref{eq:ccg}) in $d$ dimensions and retain terms of ${\cal O}(\epsilon)$
because UV counterterms appear in multiplicative renormalization.
We have
\begin{equation}
d\sigma_0=\frac{1}{2s}d{\rm PS}_2(k_1+k_2;p,k_3)\overline{|{\cal T}_0|^2},
\end{equation}
where the first factor on the right-hand side stems from the flux,
${\cal T}_0$ is the LO transition-matrix ($T$) element of process
(\ref{eq:ccg}), and it is averaged (summed) over the spin and colour states of
the incoming (outgoing) particles.
Here and in the following, we denote the Lorentz-invariant $N$-particle
phase-space element in $d$ dimensions as
\begin{equation}
d{\rm PS}_N(P;p_1,\ldots,p_N)=\mu^{(N-1)(4-d)}
(2\pi)^d\delta^{(d)}\left(P-\sum_{i=1}^Np_i\right)
\prod_{i=1}^N\frac{d^dp_i}{(2\pi)^{d-1}}\delta\left(p_i^2-m_i^2\right)
\theta\left(p_i^0\right),
\end{equation}
where $p_i$ and $m_i$ are the four-momenta and masses of the final-state
particles and $P$ is the total four-momentum of the incoming particles.
In the case of process (\ref{eq:ccg}), we have
\begin{equation}
d{\rm PS}_2(k_1+k_2;p,k_3)=
\frac{1}{8\pi s\Gamma(1-\epsilon)}
\left(\frac{4\pi\mu^2s}{tu}\right)^\epsilon\delta(s+t+u-4m^2)dt_1du_1,
\end{equation}
where $\Gamma(x)$ is Euler's $\Gamma$
function, so that
\begin{equation}
\frac{d\sigma_0}{dt_1du_1}=\frac{1}{16\pi s^2\Gamma(1-\epsilon)}
\left(\frac{4\pi\mu^2s}{tu}\right)^\epsilon\delta(s+t+u-4m^2)
\overline{|{\cal T}_0|^2}.
\label{eq:bor}
\end{equation}
Note that the spin average factor of an incoming photon or gluon in $d$
dimensions is $1/(d-2)$.
The final result is listed in Eq.~(1) of Ref.~\cite{npb}.

We now move on to the process $e^+e^-\to e^+e^-H+X$, where $H$ denotes a
generic charmonium state.
For later use, we now also include the resolved contributions.
We select the $e^+e^-$ centre-of-momentum (CM) frame and denote the nominal
$e^+e^-$ energy by $\sqrt S$, and the transverse momentum and rapidity of the
$H$ meson by $p_T$ and $y$, respectively.
Invoking the Weizs\"acker-Williams approximation (WWA) \cite{wei} and the
factorization theorems of the QCD parton model \cite{dew} and NRQCD
\cite{bbl}, the differential cross section of $e^+e^-\to e^+e^-H+X$ can be
written as
\begin{eqnarray}
\frac{d^2\sigma(e^+e^-\to e^+e^-H+X)}{dp_T^2dy}
&=&\int dx_+f_{\gamma/e}(x_+)\int dx_-f_{\gamma/e}(x_-)
\sum_{a,b,d}\int dx_af_{a/\gamma}(x_a,M)
\nonumber\\
&&{}\times
\int dx_bf_{b/\gamma}(x_b,M)\sum_n\langle{\cal O}^H[n]\rangle
s\frac{d\sigma(ab\to c\overline{c}[n]+d)}{dt_1du_1},
\label{eq:ee}
\end{eqnarray}
where $f_{\gamma/e}(x_\pm)$ is the equivalent number of transverse photons
radiated by the initial-state positrons and electrons,
$f_{a/\gamma}(x_a,M)$ are the PDFs of the photon,
$\langle{\cal O}^H[n]\rangle$ are the MEs of the $H$ meson,
$d\sigma(ab\to c\overline{c}[n]+d)$ are the differential partonic cross 
sections,
the integrals are over the longitudinal-momentum fractions of the emitted
particles w.r.t.\ the emitting ones, and it is summed over
$a,b=\gamma,g,q,\overline{q}$ and $d=g,q,\overline{q}$.
With the definition $f_{\gamma/\gamma}(x_\gamma,M)=\delta(1-x_\gamma)$,
Eq.~(\ref{eq:ee}) accommodates the direct, single-resolved, and
double-resolved channels.
The Mandelstam variables are given by $s=x_ax_bx_+x_-S$,
$t_1=-x_+x_a\sqrt Sm_T\exp(-y)$, and $u_1=-x_-x_b\sqrt Sm_T\exp(y)$, where
$m_T=\sqrt{p_T^2+4m^2}$ is the transverse mass of the $H$ meson.
For a given value of $\sqrt S$, the accessible phase space is defined by
\begin{eqnarray}
&&0\le p_T\le \frac{S-4m^2}{2\sqrt{S}},\nonumber\\
&&|y|\le\arcosh\frac{S+4m^2}{2\sqrt Sm_T}.
\end{eqnarray}

\subsection{Virtual corrections}

\begin{figure}[ht]
\begin{center}
\epsfig{figure=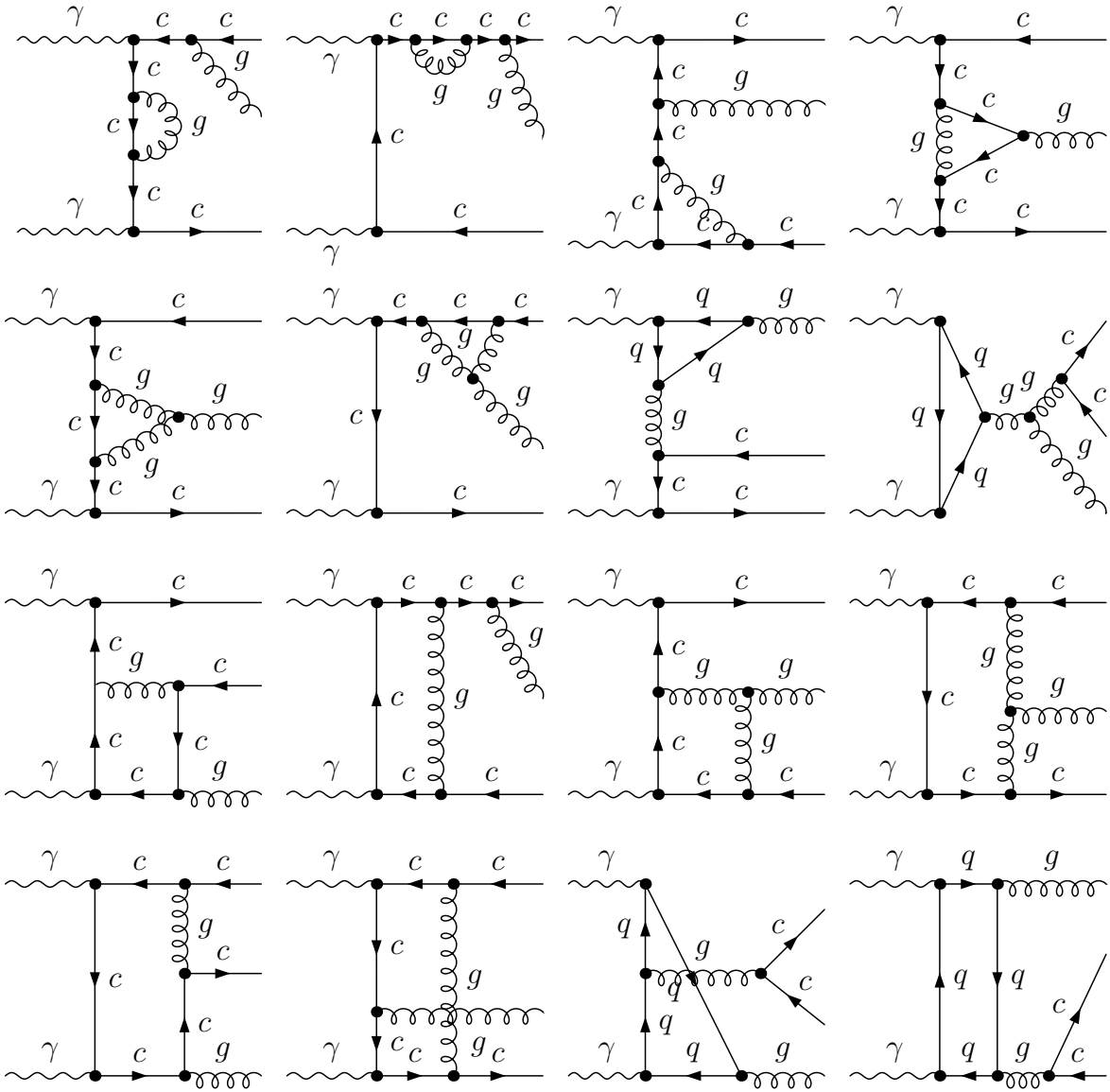,width=\textwidth,%
bbllx=74pt,bblly=214pt,bburx=512pt,bbury=632pt}
\caption{One-loop Feynman diagrams pertinent to the partonic
subprocess~(\ref{eq:ccg}).}
\label{fig:loop}
\end{center}
\end{figure}

Representative examples of diagrams that generate the virtual corrections to
the cross section of process (\ref{eq:ccg}) are shown in Fig.~\ref{fig:loop}.
They fall into two classes.
The diagrams of the first class are obtained by attaching one virtual gluon
line in all possible ways to the tree-level seed diagrams of
Fig.~\ref{fig:ccg}.
Their contribution is proportional to $e_c^2$.
They include self-energy, triangle, box, and pentagon diagrams.
Loop insertions in external gluon or charm-quark lines are accommodated in the
respective wave-function renormalization constants and are not displayed in
Fig.~\ref{fig:loop}.
The self-energy and triangle diagrams are in general UV divergent;
the triangle, box, and pentagon diagrams are in general IR divergent.
The pentagon diagrams without three-gluon vertex also contain
Coulomb singularities, which are cancelled after taking into account 
the corresponding corrections to the operator 
$\left\langle{\cal O}^H\left[{}^3\!S_1^{(8)}\right]\right\rangle$.
In the practical calculation, the Coulomb singularities are first regularized
by a small gluon mass.
This regularization prescription is then transformed into one implemented with
a small relative velocity $v$ between the $c$ and $\overline{c}$ quarks
\cite{kra}. 

The diagrams of the second class contain a light-quark triangle or box.
The triangle diagrams produce contributions proportional to $e_q^2$ or 
$e_qe_c$, which vanish individually by Furry's theorem \cite{fur}.
The box diagrams are proportional to $e_q^2$.
They contain UV and IR singularities, but their sum is finite.

Calling the $T$-matrix element comprising all these loop diagrams
${\cal T}_v$, the virtual corrections to the cross section of process
(\ref{eq:ccg}) may be evaluated as
\begin{equation}
\frac{d\sigma_v}{dt_1du_1}=\frac{1}{16\pi s^2\Gamma(1-\epsilon)}
\left(\frac{4\pi\mu^2s}{tu}\right)^\epsilon\delta(s+t+u-4m^2)
2\overline{\re\left({\cal T}_0^*{\cal T}_v\right)}.
\label{eq:vir}
\end{equation}

We apply two independent approaches to calculate the one-loop diagrams.
The first one uses {\tt FeynArts}~\cite{feynarts} to generate the diagrams and
self-written {\tt Mathematica} codes to apply the projectors and provide
expressions, which are afterwards treated with a {\tt FORM} program to perform
the tensor reduction and the extraction of the UV and IR singularities.
The result is then transformed into a {\tt Fortran} code to be used for the
numerical evaluation.
The second approach utilizes {\tt QGRAF}~\cite{Nogueira:1993ex} for the
generation of the diagrams, {\tt FeynCalc}~\cite{Mertig:an} for the tensor
reduction, and {\tt LoopTools}~\cite{looptools} for the numerical evaluation of
the IR-safe integrals.
These packages are surrounded by self-written interface programs, which allow
for a completely automated computation~\cite{Harlander:1998dq}.
As in the first approach, the relevant IR-divergent $n$-point functions are
implemented in a {\tt Fortran} code.
They can be extracted from Refs.~\cite{kra,Bern:1993kr,Beenakker:2002nc}.
For the analytical treatment of the abelian five-point functions, we refer
to Ref.~\cite{Beenakker:2002nc}.
The pentagon diagrams involving a nonabelian coupling can all be reduced to
integrals with a lower number of external legs by partial fractioning.

Our analytic result is too lengthy to be presented here.
However, the interference of the light-quark box amplitude ${\cal T}_q$ with
${\cal T}_0$ is sufficiently compact to be listed, which is done in
Eq.~(\ref{eq:lq}).

\subsubsection{Parameter and wave-function renormalization}

The self-energy and triangle diagrams of the type indicated in
Fig.~\ref{fig:loop} contain UV singularities, which are cancelled upon the
renormalizations of the QCD gauge coupling $g_s=\sqrt{4\pi\alpha_s}$, the
charm-quark mass $m$ and field $\psi$, and the gluon field $A_\mu$.
Specifically, the renormalization transformations read
\begin{equation}
g_s^0=Z_gg_s,\qquad
m^0=Z_mm,\qquad
\psi^0=\sqrt{Z_2}\psi,\qquad
A_\mu^0=\sqrt{Z_3}A_\mu,
\end{equation}
where the superscript 0 labels bare quantities and $Z_i=1+\delta Z_i$, with
$i=g,m,2,3$, are renormalization constants.
The quantities $\delta Z_i$ are of ${\cal O}(\alpha_s)$ and contain UV
singularities and, in general, also finite pieces.
The UV singularities are unique, while the finite pieces depend on the choice
of renormalization scheme.
In order to comply with the Lehmann-Symanzik-Zimmermann \cite{lsz} reduction
formula without any additional finite adjustments, we define $Z_2$ and $Z_3$
in the on-mass-shell (OS) scheme, the results being
\begin{eqnarray}
\delta Z_2^{\rm OS}&=&-C_F\frac{\alpha_s}{4\pi}
\left[\frac{1}{\epsilon_{\rm UV}}+\frac{2}{\epsilon_{\rm IR}}
-3\gamma_E+3\ln\frac{4\pi\mu^2}{m^2}+4+{\cal O}(\epsilon)\right],
\nonumber\\
\delta Z_3^{\rm OS}&=&\frac{\alpha_s}{4\pi}
\left[(\beta_0-2C_A)
\left(\frac{1}{\epsilon_{\rm UV}}-\frac{1}{\epsilon_{\rm IR}}\right)
+{\cal O}(\epsilon)\right],
\end{eqnarray}
where $\gamma_E$ is Euler's constant and $\beta_0=(11/3)C_A-(4/3)T_Fn_f$ is
the one-loop coefficient of the QCD beta function.
We also define $m$ in the OS scheme, by setting
\begin{equation}
\delta Z_m^{\rm OS}=-3C_F\frac{\alpha_s}{4\pi}
\left[\frac{1}{\epsilon_{\rm UV}}-\gamma_E+\ln\frac{4\pi\mu^2}{m^2}
+\frac{4}{3}+{\cal O}(\epsilon)\right].
\end{equation}
However, we adopt the modified minimal-subtraction ($\overline{\rm MS}$)
scheme to define $g_s$, by putting
\begin{equation}
  \delta Z_g^{\overline{\rm MS}}=-\frac{\beta_0}{2}\,
  \frac{\alpha_s}{4\pi}
  \left[\frac{1}{\epsilon_{\rm UV}} -\gamma_E + \ln(4\pi)
  \right].
\end{equation}

\subsubsection{Operator renormalization}

A crucial feature of effective field theories, such as NRQCD, is that the
composite operators are generally subject to renormalization.
In the case of NRQCD, this is essential in order to ensure the complete
cancellation of IR and Coulomb singularities at NLO and so to overcome the 
conceptual problems of the CSM mentioned in Section~\ref{sec:one}.
To be consistent with the rest of our calculation, we also employ dimensional
regularization here.
We adopt the technique described in Refs.~\cite{bbl,gre} to directly evaluate 
the NLO corrections to the NRQCD operators.
In this way, we avoid having to match partonic cross sections evaluated in
NRQCD with their counterparts in full QCD.

In the case under consideration, we have to renormalize the CO ME
$\left\langle{\cal O}^H\left[{}^3\!S_1^{(8)}\right]\right\rangle$, which
appears at LO, in $\overline{|{\cal T}_0|^2}$.
In $d$ space-time dimensions, this ME has mass dimension $d-1$.
We thus introduce the 't~Hooft mass scale of NRQCD, $\lambda$, to keep its
renormalized version, which we wish to extract from experimental data, at mass
dimension 3.

\begin{figure}[ht]
\begin{center}
\epsfig{figure=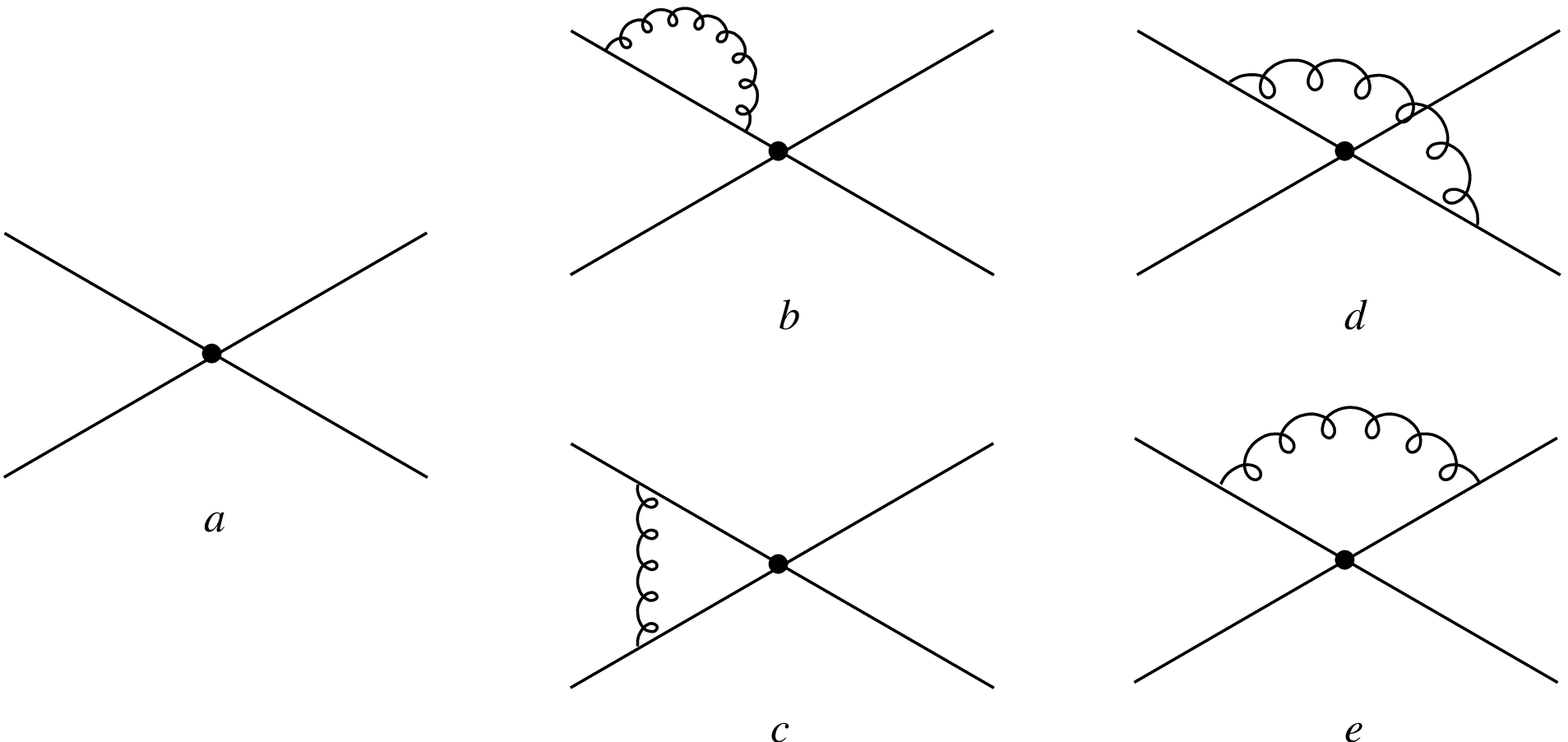,width=\textwidth}
\caption{Feynman diagrams pertinent to the one-loop correction to
$\left\langle{\cal O}^H\left[{}^3\!S_1^{(8)}\right]\right\rangle$.}
\label{fig:op}
\end{center}
\end{figure}

The four-quark operator
$\left\langle{\cal O}^H\left[{}^3\!S_1^{(8)}\right]\right\rangle$ is related
to the amplitude for the elastic scattering of a $c\overline{c}$ pair. 
The corresponding tree-level diagram is depicted in Fig.~\ref{fig:op}(a).
The one-loop corrections to this amplitude are obtained by attaching a virtual
gluon line in all possible ways to the external heavy-quark legs, and they
involve self-energy and vertex corrections [see Fig.~\ref{fig:op}(b)--(e)].
Using the NRQCD Feynman rules in the quarkonium rest frame, expanding the
one-loop integrands as Taylor series in $1/m$, and performing the integration
over the loop momentum, we obtain the unrenormalized one-loop result
\begin{eqnarray}
\left\langle{\cal O}^H\left[{}^3\!S_1^{(8)}\right]\right\rangle_1
&=&\left\langle{\cal O}^H\left[{}^3\!S_1^{(8)}\right]\right\rangle_0
\left[1+\left(C_F-\frac{C_A}{2}\right)\frac{\pi\alpha_s}{2v}\right]
+\frac{4\alpha_s}{3\pi m^2}\left(\frac{4\pi\mu^2}{\lambda^2}\right)^\epsilon
\exp(-\epsilon\gamma_E)
\nonumber\\
&&{}\times\left(\frac{1}{\epsilon_{\rm UV}}-\frac{1}{\epsilon_{\rm IR}}\right)
\sum_{J=0}^2\left(
C_F\left\langle{\cal O}^H\left[{}^3\!P_J^{(1)}\right]\right\rangle
+B_F\left\langle{\cal O}^H\left[{}^3\!P_J^{(8)}\right]\right\rangle\right),
\label{eq:reg}
\end{eqnarray}
where the subscript 0 labels the tree-level quantity and $\mu$ is the 't~Hooft
mass scale of QCD that enters through the $d$-dimensional loop integration.
The term proportional to $1/v$ represents the Coulomb singularity, which
arises from the exchange of a longitudinal gluon between the outgoing $c$ and
$\overline{c}$ quarks [see Fig.~\ref{fig:op}(c)].
Obviously, NRQCD operators of different $c\overline{c}$ Fock states $n$ start
to mix at one loop.
Furthermore, the presence of UV singularities indicates that they need
renormalization.
In the following, we choose the $\overline{\rm MS}$ scheme for that.
We thus write
\begin{eqnarray}
\left\langle{\cal O}^H\left[{}^3\!S_1^{(8)}\right]\right\rangle_1
&=&\left\langle{\cal O}^H\left[{}^3\!S_1^{(8)}\right]\right\rangle_r(\lambda)
+\frac{4\alpha_s}{3\pi m^2}
\left(\frac{4\pi\mu^2}{\lambda^2}\right)^\epsilon\exp(-\epsilon\gamma_E)
\frac{1}{\epsilon_{\rm UV}}\nonumber\\
&&{}\times\sum_{J=0}^2\left(
C_F\left\langle{\cal O}^H\left[{}^3\!P_J^{(1)}\right]\right\rangle
+B_F\left\langle{\cal O}^H\left[{}^3\!P_J^{(8)}\right]\right\rangle\right),
\label{eq:ct}
\end{eqnarray}
where the subscript $r$ labels the renormalized quantity and we identify
$\lambda$ with the NRQCD renormalization scale.
Inserting Eq.~(\ref{eq:ct}) into Eq.~(\ref{eq:reg}), we obtain
\begin{eqnarray}
\left\langle{\cal O}^H\left[{}^3\!S_1^{(8)}\right]\right\rangle_0
&=&\left\langle{\cal O}^H\left[{}^3\!S_1^{(8)}\right]\right\rangle_r(\lambda)
\left[1-\left(C_F-\frac{C_A}{2}\right)\,\frac{\pi\alpha_s}{2v}\right]
+\frac{4\alpha_s}{3\pi m^2}
\left(\frac{4\pi\mu^2}{\lambda^2}\right)^\epsilon\exp(-\epsilon\gamma_E)
\frac{1}{\epsilon_{\rm IR}}\nonumber\\
&&{}\times\sum_{J=0}^2\left(
C_F\left\langle{\cal O}^H\left[{}^3\!P_J^{(1)}\right]\right\rangle
+B_F\left\langle{\cal O}^H\left[{}^3\!P_J^{(8)}\right]\right\rangle\right).
\label{eq:ren}
\end{eqnarray}
Combining Eq.~(\ref{eq:ren}) with Eq.~(\ref{eq:bor}) generates an IR
counterterm at ${\cal O}(\alpha_s)$ that is indispensible to render the total
NLO result finite.
The Coulomb singularity present in Eq.~(\ref{eq:ren}) is necessary to cancel
similar terms in the virtual corrections.

\subsubsection{Renormalization group equation}

The renormalization group equation that determines the $\lambda$ dependence of
$\left\langle{\cal O}^H\left[{}^3\!S_1^{(8)}\right]\right\rangle_r(\lambda)$
may be derived by differentiating Eq.~(\ref{eq:ren}) with respect to $\lambda$
and then taking the physical limit $\epsilon\to0$.
It reads
\begin{equation}
\frac{\lambda^2d}{d\lambda^2}
\left\langle{\cal O}^H\left[{}^3\!S_1^{(8)}\right]\right\rangle_r(\lambda)
=\frac{4\alpha_s(\mu)}{3\pi m^2}\sum_{J=0}^2\left(
C_F\left\langle{\cal O}^H\left[{}^3\!P_J^{(1)}\right]\right\rangle
+B_F\left\langle{\cal O}^H\left[{}^3\!P_J^{(8)}\right]\right\rangle\right).
\label{eq:rge}
\end{equation}
There is no obvious physical reason to distinguish between the scales $\mu$
and $\lambda$, which refer to the $c\overline{c}g$ and
$c\overline{c}c\overline{c}$ vertices in the one-loop diagrams of
Figs.~\ref{fig:op}(b)--(e), respectively.
Both scales should essentially be of ${\cal O}(m)$.
In the following, we thus identify $\mu=\lambda$.
Integration of Eq.~(\ref{eq:rge}) then yields
\begin{eqnarray}
\left\langle{\cal O}^H\left[{}^3\!S_1^{(8)}\right]\right\rangle_r(\lambda)
&=&\left\langle{\cal 
O}^H\left[{}^3\!S_1^{(8)}\right]\right\rangle_r(\lambda_0)
+\frac{16}{3m^2}\sum_{J=0}^2\left(
C_F\left\langle{\cal O}^H\left[{}^3\!P_J^{(1)}\right]\right\rangle
+B_F\left\langle{\cal O}^H\left[{}^3\!P_J^{(8)}\right]\right\rangle\right)
\nonumber\\
&&{}\times\left\{\frac{1}{\beta_0}
  \ln\frac{\alpha_s(\lambda_0)}{\alpha_s(\lambda)}
  +\frac{\beta_1}{4\pi\beta_0^2}
  \left[\alpha_s(\lambda)-\alpha_s(\lambda_0)\right]
\right\},
\label{eq::oprun}
\end{eqnarray}
where $\lambda_0$ is a reference scale for which the value of 
$\left\langle{\cal O}^H\left[{}^3\!S_1^{(8)}\right]\right\rangle_r(\lambda_0)$
is assumed to be known and $\beta_1=(34/3)C_A^2-4C_FT_Fn_f-(20/3)C_AT_Fn_f$ is
the two-loop coefficient of the QCD beta function.
In want of a genuine NLO determination of
$\left\langle{\cal O}^H\left[{}^3\!S_1^{(8)}\right]\right\rangle_r(\lambda_0)$
from a fit to experimental data, we choose $\lambda_0=m$ and identify
$\left\langle{\cal O}^H\left[{}^3\!S_1^{(8)}\right]\right\rangle_r(m)$ with
its $\lambda$-independent LO value, which is known from the literature
\cite{bkl}.
Since we consider the NLO corrections to the LO cross section~(\ref{eq:bor}),
which is already of ${\cal O}(\alpha_s)$, we have to employ in 
Eq.~(\ref{eq::oprun}) the two-loop
formula for $\alpha_s(\lambda)$, which reads
\begin{equation}
\frac{\alpha_s(\lambda)}{4\pi}=\frac{1}{\beta_0L}
-\frac{\beta_1\ln L}{\beta_0^3L^2},
\label{eq:as}
\end{equation}
where $L=\ln\left(\lambda^2/\Lambda_{\rm QCD}^2\right)$.

\subsection{Real corrections}

\begin{figure}[ht]
\begin{center}
\begin{tabular}{c}
\parbox{\textwidth}{\epsfig{figure=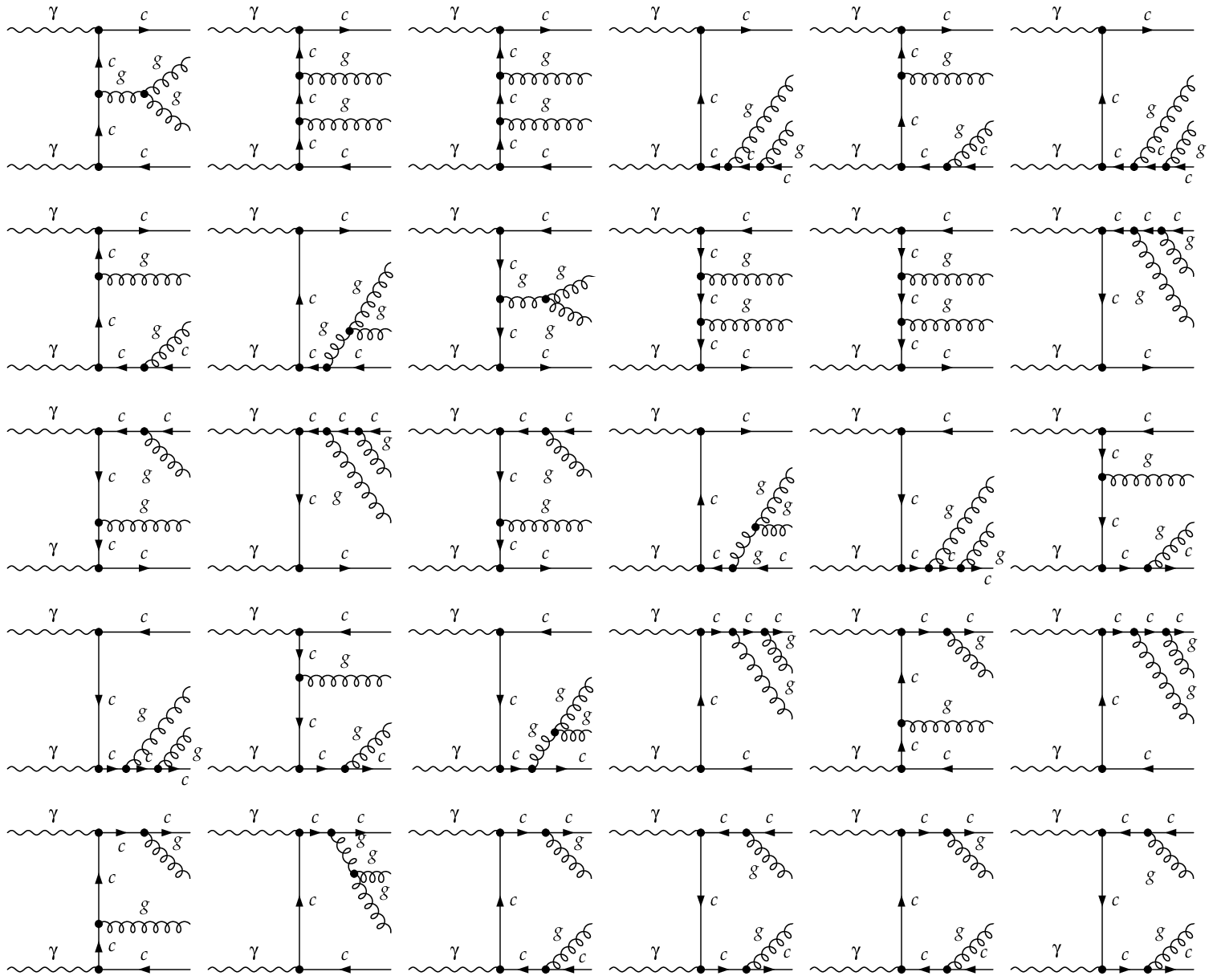,width=\textwidth}} \\
(a) \\
\parbox{\textwidth}{\epsfig{figure=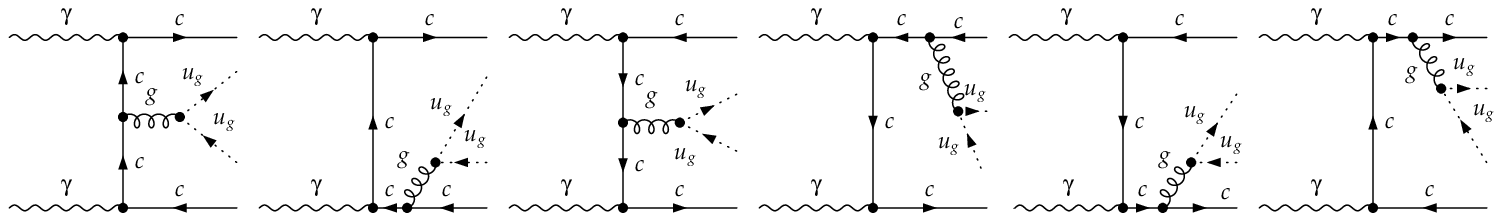,width=\textwidth,
bbllx=0pt,bblly=385pt,bburx=432pt,bbury=485pt}} \\
(b)
\end{tabular}
\caption{Tree-level Feynman diagrams pertinent to the partonic subprocesses (a)
(\ref{eq:ccgg}) and (b) (\ref{eq:ccuu}).}
\label{fig:ccgg}
\end{center}
\end{figure}

\begin{figure}[ht]
\begin{center}
\epsfig{figure=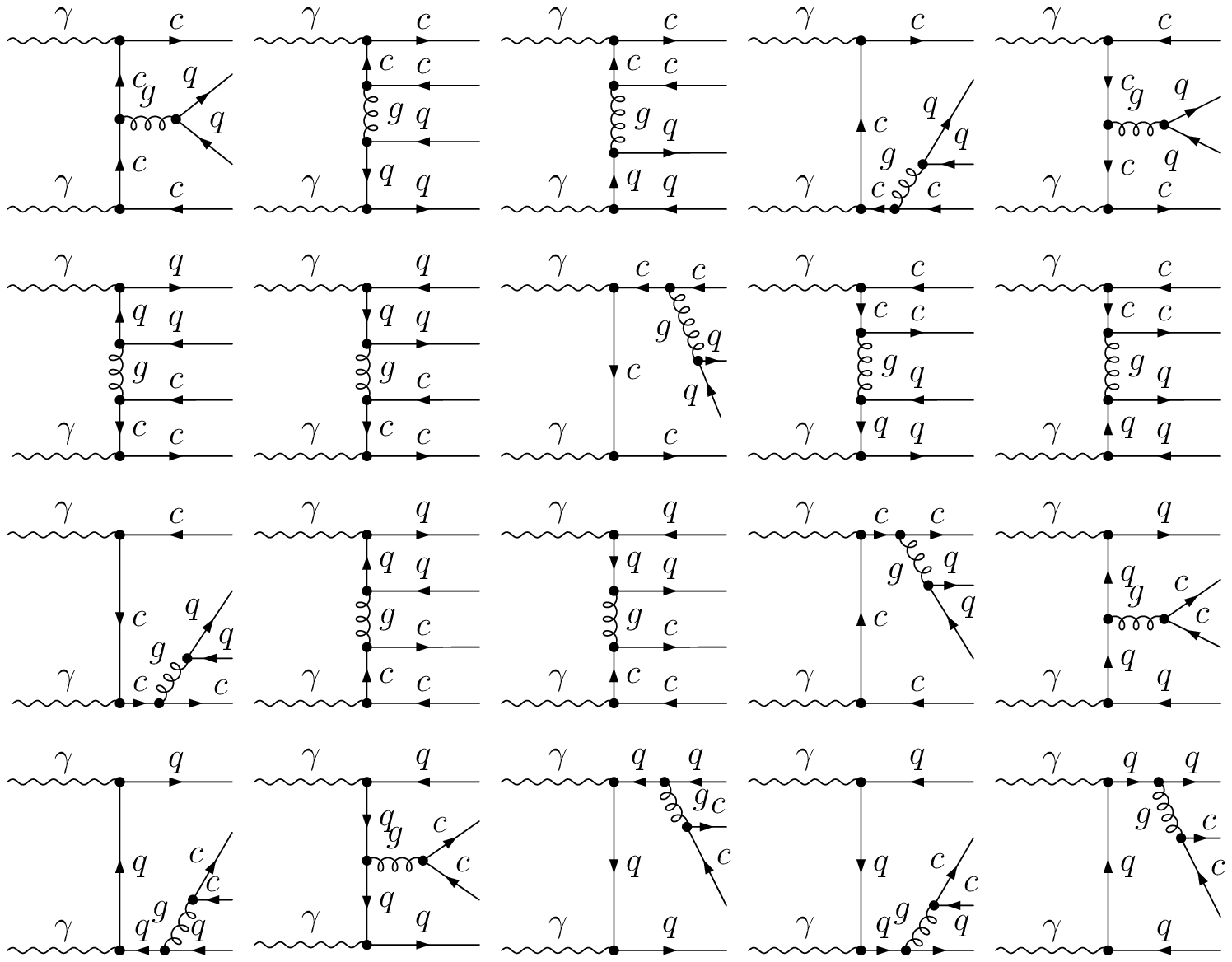,width=\textwidth,
bbllx=74pt,bblly=274pt,bburx=512pt,bbury=632pt}
\caption{Tree-level Feynman diagrams pertinent to the partonic
subprocess~(\ref{eq:ccqq}).}
\label{fig:ccqq}
\end{center}
\end{figure}

The real corrections to the cross section of process (\ref{eq:ccg}) arise from
the partonic subprocesses
\begin{equation}
\gamma(k_1)+\gamma(k_2)\to c\overline{c}[n](p)+g(k_3)+g(k_4),
\label{eq:ccgg}
\end{equation}
where $n={}^3\!P_J^{(1)},{}^1\!S_0^{(8)},{}^3\!S_1^{(8)},{}^3\!P_J^{(8)}$, and
\begin{equation}
\gamma(k_1)+\gamma(k_2)\to c\overline{c}[n](p)+q(k_3)+\overline{q}(k_4),
\label{eq:ccqq}
\end{equation}
where $n={}^1\!S_0^{(8)},{}^3\!S_1^{(8)},{}^3\!P_J^{(8)}$.
The respective diagrams are presented in Figs.~\ref{fig:ccgg}(a) and
\ref{fig:ccqq}.
From the technical point of view, it is convenient to evaluate the gluon
polarization sum as
$\sum_{\rm pol}\varepsilon_\mu^a(q)\varepsilon_\nu^{b*}(q)
=-\delta^{ab}g_{\mu\nu}$,
at the expense of allowing for the Faddeev-Popov ghosts of the gluon, $u_g$
and $\overline{u}_g$, to appear in the final state.
The additional partonic subprocesses are
\begin{equation}
\gamma(k_1)+\gamma(k_2)\to c\overline{c}[n](p)+u_g(k_3)+\overline{u}_g(k_4),
\label{eq:ccuu}
\end{equation}
where $n={}^3\!S_1^{(8)}$.
The corresponding diagrams are depicted in Fig.~\ref{fig:ccgg}(b).

Process (\ref{eq:ccgg}) with $n={}^3\!S_1^{(1)}$ is forbidden by Furry's
theorem \cite{fur}, as may be understood by observing that the
${}^3\!S_1^{(1)}$ projector effectively closes the charm-quark line and acts
like a vector coupling and that the two gluons are then in a CS state, so that
we are dealing with a closed fermion loop containing five vector couplings.
This was also verified by explicit calculation.
On the other hand, process (\ref{eq:ccqq}) with
$n={}^3\!S_1^{(1)},{}^3\!P_J^{(1)}$ is prohibited by colour conservation,
since the two quark lines are connected by a single gluon, which ensures that
the $c\bar c$ and $q\bar q$ pairs are both in a CO state.

The diagrams in Fig.~\ref{fig:ccqq} can be divided into three classes.
The diagrams of the first (second) class are proportional to $e_q^2$
($e_c^2$) and contain a timelike virtual gluon that splits into a
$c\overline{c}$ ($q\overline{q}$) pair.
The diagrams of the third class are proportional to $e_qe_c$ and contain a
spacelike virtual gluon.

As for the kinematics of processes (\ref{eq:ccgg})--(\ref{eq:ccuu}), we now
have $k_1+k_2=p+k_3+k_4$, $k_1^2=k_2^2=k_3^2=k_4^2=0$, and $p^2=4m^2$.
We can define $\left({5\atop2}\right)=10$ Mandelstam variables, 5 of which are
linearly independent.
In addition to $s_1$, $t_1$, and $u_1$ defined in Eq.~(\ref{eq:stu}), we
introduce
\begin{eqnarray}
s_3&=&(p+k_3)^2-4m^2,\qquad
s_4=(p+k_4)^2-4m^2,\qquad
s_{34}=(k_3+k_4)^2,\nonumber\\
s_{13}&=&(k_1-k_3)^2,\qquad
s_{23}=(k_2-k_3)^2,\qquad
s_{14}=(k_1-k_4)^2,\qquad
s_{24}=(k_2-k_4)^2.
\end{eqnarray}
If we take $s_1$, $t_1$, $u_1$, $s_{14}$, and $s_{24}$ to be the set of
independent Mandelstam variables, then the others are given by
\begin{eqnarray}
s_3&=&s_1+s_{14}+s_{24},\nonumber\\
s_4&=&-s_1-t_1-u_1-s_{14}-s_{24}-8m^2,\nonumber\\
s_{34}&=&s_1+t_1+u_1+8m^2,\nonumber\\
s_{13}&=&-s_1-t_1-s_{14}-4m^2,\nonumber\\
s_{23}&=&-s_1-u_1-s_{24}-4m^2.
\end{eqnarray}

Generically denoting the $T$-matrix elements of processes
(\ref{eq:ccgg})--(\ref{eq:ccuu}) by ${\cal T}_r$, their cross sections may be
evaluated as
\begin{equation}
d\sigma_r=\frac{1}{2s}d{\rm PS}_3(k_1+k_2;p,k_3,k_4)\overline{|{\cal T}_r|^2}.
\label{eq:xsr}
\end{equation}
In the case of process~(\ref{eq:ccgg}), a factor of 1/2 has to be included on
the right-hand side of Eq.~(\ref{eq:xsr}) to account for the fact that there
are two identical particles in the final state.

Integrating Eq.~(\ref{eq:xsr}) over the three-particle phase space while
keeping the value of $p_T$ finite, we encounter IR singularities, which can be
of the soft and/or collinear type.
In order to systematically extract these singularities in a Lorentz-invariant
way, it is useful to slice the phase space by introducing infinitesimal
dimensionless cut-off parameters $\delta_i$ and $\delta_f$, which are
connected with the initial and final states, respectively \cite{harris}.
In the case of processes~(\ref{eq:ccgg}) and (\ref{eq:ccuu}), we are thus led
to distinguish between the following regions of phase space.
In the {\it soft} region, where either $s_3+s_{34}<\delta_fs$ or
$s_4+s_{34}<\delta_fs$, one of the outgoing gluons is soft and, in addition,
they may be collinear.
In the {\it final-state collinear} region, where 
$s_{34}<\delta_fs<s_3+s_{34},s_4+s_{34}$, the outgoing gluons are both hard
and collinear.
In the {\it hard} region, which comprises the residual phase space, none of
the outgoing gluons is soft and they are not collinear.
In the case of process~(\ref{eq:ccqq}), we differentiate between the following
regions of phase space.
In the {\it initial-state collinear} region, where one of the inequalities
$s_{13}<\delta_is$, $s_{14}<\delta_is$, $s_{23}<\delta_is$, and
$s_{24}<\delta_is$ is satisfied, one of the outgoing $q$ and $\overline{q}$
quarks is collinear to one of the incoming photons.
In the {\it final-state collinear} region, where 
$s_{34}<\delta_fs<s_3+s_{34},s_4+s_{34}$, the outgoing $q$ and $\overline{q}$
quarks are both hard and collinear.
In the {\it hard} region, which includes the left-over phase space, none of
the outgoing $q$ and $\overline{q}$ quarks is soft and they are not collinear.
While the various contributions depend on the cut-off parameters $\delta_i$
and/or $\delta_f$, their sum must be independent of them.
The verification of the cut-off independence provides a nontrivial and 
powerful
check for the correctness of our calculation.

\subsubsection{Hard region}

The integration of Eq.~(\ref{eq:xsr}) over the hard region can be facilitated
by decomposing the three-particle phase space as
\begin{equation}
d{\rm PS}_3(k_1+k_2;p,k_3,k_4)=\frac{1}{2\pi}
d{\rm PS}_2^\star(k_1+k_2;p,k_{34})d{\rm PS}_2(k_{34};k_3,k_4),
\end{equation}
where $k_{34}=k_3+k_4$ and
\begin{eqnarray}
d{\rm PS}_2^\star(k_1+k_2;p,k_{34})&=&\mu^{4-d}
\frac{d^dp}{(2\pi)^{d-1}}\delta(p^2-4m^2)\theta(p^0)
\frac{d^dk_{34}}{(2\pi)^{d-1}}(2\pi)^d\delta^{(d)}(k_1+k_2-p-k_{34})
\nonumber\\
&=&\frac{1}{8\pi s\Gamma(1-\epsilon)}
\left(\frac{4\pi\mu^2s}{t_1u_1-4m^2s}\right)^\epsilon dt_1du_1.
\end{eqnarray}
On the other hand, writing
\begin{eqnarray}
p^\mu&=&\left(p^0,\vec{0}_T,p\right),\nonumber\\
k_3^\mu&=&\left(k_3^0,\ldots,k_3^0\sin\theta_3\sin\phi_3,
k_3^0\sin\theta_3\cos\phi_3,k_3^0\cos\theta_3\right),\nonumber\\
k_4^\mu&=&\left(k_4^0,\ldots,k_4^0\sin\theta_4\sin\phi_4,
k_4^0\sin\theta_4\cos\phi_4,k_4^0\cos\theta_4\right),
\end{eqnarray}
we have
\begin{eqnarray}
d{\rm PS}_2(k_{34};k_3,k_4)
&=&\frac{\Gamma(1-\epsilon)}{16\pi^2\Gamma(1-2\epsilon)}
\left(\frac{4\pi\mu^2}{\left(k_3^0\right)^2}\right)^\epsilon
\frac{s+4m^2-2\sqrt sp^0}{(\sqrt s-p^0+2p\cos\theta_3)^2}
\nonumber\\
&&{}\times
\sin^{d-3}\theta_3d\theta_3\sin^{d-4}\phi_3d\phi_3,
\end{eqnarray}
where the boundaries of integration are fixed by the kinematical constraints
specified above.
Since the contribution from the hard region is UV and IR finite, we may
perform the phase-space integration in $d=4$ space-time dimensions.

\subsubsection{Soft region}

As an example, let us consider the kinematical situation where $k_3$ in
processes~(\ref{eq:ccgg}) or (\ref{eq:ccuu}) becomes sufficiently soft, so
that it does not influence the kinematics of the hard process involving the
residual four-momenta.
In practice, this is arranged by imposing the condition $s_3+s_{34}<\delta_fs$
with a sufficiently small value of $\delta_f$.
The three-particle phase space can then be decomposed as
\begin{equation}
d{\rm PS}_3(k_1+k_2;p,k_3,k_4)=d{\rm PS}_2(k_1+k_2;p,k_4)d{\rm PS}_s^3,
\end{equation}
where
\begin{eqnarray}
d{\rm PS}_s^3&=&\mu^{4-d}\frac{d^{d-1}k_3}{(2\pi)^{d-1}2k_3^0}\nonumber\\
&=&\frac{\Gamma(1-\epsilon)}{8\pi^3\Gamma(1-2\epsilon)}
\left(\frac{\pi\mu^2}{\left(k_3^0\right)^2}\right)^\epsilon
k_3^0dk_3^0\sin^{d-3}\theta_3d\theta_3\sin^{d-4}\phi_3d\phi_3.
\end{eqnarray}
At the same time, $\overline{|{\cal T}_r|^2}$ factorizes as
\begin{equation}
\overline{|{\cal T}_r|^2}=\overline{|{\cal T}_0|^2}F_3,
\label{eq:soft}
\end{equation}
where $F_3$ is an appropriate Eikonal factor, the specific form of which
depends on the $c\overline{c}$ Fock state $n$.
Specifically, soft singularities occur for processes~(\ref{eq:ccgg}) and
(\ref{eq:ccuu}) with $n={}^3\!P_J^{(1)},{}^3\!S_1^{(8)},{}^3\!P_J^{(8)}$.
The appropriate expressions for $F_3$ are listed in Table~\ref{tab:f}.
Notice that, in Eq.~(\ref{eq:soft}), ${\cal T}_0$ always refers to
process~(\ref{eq:ccg}) with $n={}^3\!S_1^{(8)}$.

\begin{table}[t]
\begin{center}
\begin{tabular}{|l|l|l|}
\hline\hline
Process & Limit & Eikonal factor \\
\hline
$\gamma\gamma\to c\overline{c}\left[{}^3\!P_J^{(1)}\right]gg$ &
$k_3\to0$ &
$F_3=\frac{16C_Fg_s^2}{3s_3^2}\left(1-\frac{\epsilon}{3}\right)$ \\
& $k_4\to0$ &
$F_4=\frac{16C_Fg_s^2}{3s_4^2}\left(1-\frac{\epsilon}{3}\right)$ \\
$\gamma\gamma\to c\overline{c}\left[{}^3\!S_1^{(8)}\right]gg$ &
$k_3\to0$ &
$F_3=\frac{4C_Ag_s^2}{s_3}\left(\frac{s_4}{s_{34}}-\frac{4m^2}{s_3}\right)$ \\
& $k_4\to 0$ &
$F_4=\frac{4C_Ag_s^2}{s_4}\left(\frac{s_3}{s_{34}}-\frac{4m^2}{s_4}\right)$ \\
& $k_3\parallel k_4$ &
$F_{34}=\frac{4C_Ag_s^2}{s_{34}}
\left[\frac{z}{1-z}+\frac{1-z}{z}+z(1-z)\right]$ \\
$\gamma\gamma\to c\overline{c}\left[{}^3\!P_J^{(8)}\right]gg$ &
$k_3\to0$ &
$F_3=\frac{16B_Fg_s^2}{3s_3^2}\left(1-\frac{\epsilon}{3}\right)$ \\
& $k_4\to 0$ &
$F_4=\frac{16B_Fg_s^2}{3s_4^2}\left(1-\frac{\epsilon}{3}\right)$ \\
$\gamma\gamma\to c\overline{c}\left[{}^3\!S_1^{(8)}\right]q\bar{q}$ &
$k_3\parallel k_4$ &
$F_{34}=\frac{2T_Fg_s^2}{s_{34}}\left[z^2+(1-z)^2-2\epsilon z(1-z)\right]$ \\
$\gamma\gamma\to c\overline{c}[n]q\overline{q}$ &
$k_2\parallel k_3$ &
$F_{23}=-\frac{2e_q^2e^2}{s_{23}}\,\frac{z^2+(1-z)^2-\epsilon}{z}$ \\
& $k_2\parallel k_4$ &
$F_{24}=-\frac{2e_q^2e^2}{s_{24}}\,\frac{z^2+(1-z)^2-\epsilon}{z}$ \\
\hline\hline
\end{tabular}
\caption{Eikonal factors appropriate for the soft and collinear limits of the
squared $T$-matrix elements of the various partonic subprocesses.
Here, $n={}^1\!S_0^{(8)},{}^3\!S_1^{(8)},{}^3\!P_J^{(8)}$.}
\label{tab:f}
\end{center}
\end{table}

The integration of the Eikonal factors over the soft parts of the phase space 
can be performed analytically, and the results for the soft integrals,
\begin{equation}
I_s^3=\int_{k_3^0<\delta_f\sqrt s/2}d{\rm PS}_s^3F_3
\end{equation}
and similarly for $k_4$, may be found in Table~\ref{tab:i}, where
\begin{eqnarray}
C_\epsilon&=&\frac{1}{(4\pi)^2}
\left(\frac{4\pi\mu^2}{m^2}\right)^\epsilon\exp(-\epsilon\gamma_E),
\nonumber\\
\beta&=&\frac{s-4m^2}{s+4m^2}.
\label{eq::def_ceps}
\end{eqnarray}

\begin{table}[t]
\begin{center}
\begin{tabular}{|l|l|}
\hline\hline
Process & Soft and collinear integrals \\
\hline
$\gamma\gamma\to c\overline{c}\left[{}^3\!P_J^{(1)}\right]gg$ &
$I_s^3=-\frac{8C_Fg_s^2}{3m^2}C_{\epsilon}
\left(\frac{m^2}{\delta_f^2s}\right)^\epsilon
\left(\frac{1}{\epsilon}
+\frac{1}{\beta}\ln\frac{1+\beta}{1-\beta}-\frac{1}{3}\right)$ \\
& $I_s^4=I_s^3$ \\
$\gamma\gamma\to c\overline{c}\left[{}^3\!S_1^{(8)}\right]gg$ &
$I_s^3=2C_Ag_s^2C_\epsilon\left(\frac{m^2}{\delta_f^2s}\right)^\epsilon
\left[\frac{1}{\epsilon^2}-\frac{1}{\epsilon}\left(
\ln\frac{1+\beta}{1-\beta}-1\right)
-2\li\left(\frac{2\beta}{1-\beta}\right)
\right.$ \\
& $\hphantom{I_s^3={}}
-\left.\frac{1}{2}\ln^2\frac{1+\beta}{1-\beta}
+\frac{1}{\beta}\ln\frac{1+\beta}{1-\beta}-\frac{\pi^2}{4}\right]$ \\
& $I_c^{34}=-2C_Ag_s^2C_{\epsilon}\left(\frac{m^2}{\delta_fs}\right)^\epsilon
\left[\frac{1}{\epsilon}
\left(2\ln\frac{s_1}{\delta_fs}-\frac{11}{6}\right)
+\ln^2\frac{s_1}{\delta_fs}+\frac{\pi^2}{3}-\frac{67}{18}\right]$ \\
$\gamma\gamma\to c\overline{c}\left[{}^3\!P_J^{(8)}\right]gg$ &
$I_s^3=-\frac{8B_Fg_s^2}{3m^2}C_\epsilon
\left(\frac{m^2}{\delta_f^2s}\right)^\epsilon
\left(\frac{1}{\epsilon}
+\frac{1}{\beta}\ln\frac{1+\beta}{1-\beta}-\frac{1}{3}\right)$ \\
& $I_s^4=I_s^3$ \\
$\gamma\gamma\to c\overline{c}\left[{}^3\!S_1^{(8)}\right]q\bar{q}$ &
$I_c^{34}=-\frac{4}{3}T_fn_fg_s^2C_{\epsilon}
\left(\frac{m^2}{\delta_fs}\right)^\epsilon
\left(\frac{1}{\epsilon}+\frac{5}{3}\right)$ \\
\hline\hline
\end{tabular}
\caption{Soft and collinear integrals.}
\label{tab:i}
\end{center}
\end{table}

\subsubsection{Final-state collinear region}

In the final-state collinear region, where
$s_{34}<\delta_fs<s_3+s_{34},s_4+s_{34}$, it is useful to introduce the
parameterization
\begin{eqnarray}
k_3^\mu&=&\left(zp+\frac{k_T^2}{2zp},\vec{k}_T,-zp\right),\nonumber\\
k_4^\mu&=&\left((1-z)p+\frac{k_T^2}{2(1-z)p},-\vec{k_T},-(1-z)p\right),
\end{eqnarray}
where $z$, with $0<z<1$, determines the fractions of longitudinal momentum
that the final-state partons receive from the splitting gluon,
$k_T^2=z(1-z)s_{34}$, and terms of ${\cal O}\left(k_T^4/p^4\right)$ are
neglected.
We can then decompose the three-particle phase space as
\begin{equation}
d{\rm PS}_3(k_1+k_2;p,k_3,k_4)=d{\rm PS}_2(k_1+k_2;p,k_{34})d{\rm PS}_c^{34},
\end{equation}
where
\begin{eqnarray}
d{\rm PS}_c^{34}&=&\mu^{4-d}\frac{d^{d-1}k_3}{(2\pi)^{d-1}2k_3^0}\,
\frac{k_{34}^0}{k_4^0}\nonumber\\
&=&\frac{1}{16\pi^2\Gamma(1-\epsilon)}
\left(\frac{4\pi\mu^2}{z(1-z)s_{34}}\right)^\epsilon dzds_{34}.
\end{eqnarray}
Using the relation
\begin{equation}
\frac{1}{\Gamma(1-\epsilon)}=\frac{\Gamma(1-\epsilon)}{\Gamma(1-2\epsilon)}
+{\cal O}(\epsilon^2),
\end{equation}
where the terms of ${\cal O}(\epsilon^2)$ are inconsequential in the physical
limit $d\to4$ because, in the collinear limit, $\overline{|{\cal T}_r|^2}$
only develops simple poles in $\epsilon$, we recover the familiar combination
of Euler's Gamma functions.
At the same time, $\overline{|{\cal T}_r|^2}$ factorizes as
\begin{equation}
\overline{|{\cal T}_r|^2}=\overline{|{\cal T}_0|^2}F_{34},
\end{equation}
where $F_{34}$ is an appropriate eikonal factor.
In the case under consideration, final-state collinear singularities occur for
processes~(\ref{eq:ccgg}) and (\ref{eq:ccqq}) with $n={}^3\!S_1^{(8)}$, where
they are related to $g\to gg$ and $g\to q\overline{q}$ splitting,
respectively.
The corresponding expressions for $F_{34}$ may be found in Table~\ref{tab:f}.
They are related to the timelike $g\to g$ and $g\to q$ splitting functions,
\begin{eqnarray}
P_{g\to g}(z)&=&2C_A\left[\frac{z}{1-z}+\frac{1-z}{z}+z(1-z)\right],
\\
P_{g\to q}(z)&=&T_F\left[z^2+(1-z)^2\right],
\label{eq:pgq}
\end{eqnarray}
as
\begin{eqnarray}
F_{34}&=&\frac{2g_s^2}{s_{34}}P_{g\to g}(z),
\nonumber\\
F_{34}&=&\frac{2g_s^2}{s_{34}}\,\left[P_{g\to q}(z)-2\epsilon T_F z(1-z)\right],
\end{eqnarray}
respectively.

The integration of $F_{34}$ over the collinear parts of the phase space can be
performed analytically, and the resulting collinear integrals,
\begin{equation}
I_c^{34}=\int_{s_{34}<\delta_f s}d{\rm PS}_c^{34}F_{34},
\label{eq:coll}
\end{equation}
may be found in Table~\ref{tab:i}.
In the case of process~(\ref{eq:ccgg}), the range of integration in
Eq.~(\ref{eq:coll}) has to be further constrained by the condition
$\delta_fs/s_1<z<1-\delta_fs/s_1$ to ensure that the final-state gluons are
both hard.

\subsubsection{Initial-state collinear region}

As an example, let us consider the kinematical situation where the incoming
photon with four-momentum $k_2$ in process~(\ref{eq:ccqq}) splits into a hard,
collinear, quasireal antiquark $\overline q$, with four-momentum $k_2^\prime$,
and the outgoing quark $q$, with four-momentum $k_3$.
The collinearity is enforced by the constraint $s_{23}<\delta_is$ with a
sufficiently small value of $\delta_i$.
It is then helpful to introduce the parameterization
\begin{eqnarray}
k_2^\mu&=&\left(k_2,\vec{0}_T,k_2\right),\nonumber\\
k_2^{\prime\mu}&=&\left(zk_2+\frac{k_T^2}{2zk_2},\vec{k}_T,zk_2\right),
\nonumber\\
k_3^\mu&=&\left((1-z)k_2+\frac{k_T^2}{2(1-z)k_2},-\vec{k}_T,(1-z)k_2\right),
\end{eqnarray}
where $k_T^2=-(1-z)s_{23}$ and terms of ${\cal O}\left(k_T^4/k_2^4\right)$ are
neglected.
We can then decompose the three-particle phase space as
\begin{equation}
d{\rm PS}_3(k_1+k_2;p,k_3,k_4)=d{\rm PS}_2(k_1+zk_2;p,k_4)d{\rm PS}_c^{23},
\end{equation}
where
\begin{eqnarray}
d{\rm PS}_c^{23}&=&\mu^{4-d}\frac{d^{d-1}k_3}{(2\pi)^{d-1}2k_3^0}\nonumber\\
&=&\frac{1}{16\pi^2\Gamma(1-\epsilon)}
\left(\frac{4\pi\mu^2}{-(1-z)s_{23}}\right)^\epsilon dzds_{23}.
\end{eqnarray}
At the same time, $\overline{|{\cal T}_r|^2}$ factorizes as
\begin{equation}
\overline{|{\cal T}_r|^2}=\overline{|{\cal T}_0|^2}\left(\gamma(k_1)
+\overline{q}\left(k_2^\prime\right)\to c\overline{c}[n](p)+\overline{q}(k_4)
\right)F_{23},
\label{eq:ini}
\end{equation}
where the $c\overline{c}$ Fock state $n$ is the same as appears in
${\cal T}_r$ and $F_{23}$ is an appropriate Eikonal factor.
In the case under consideration, initial-state collinear singularities occur
for process~(\ref{eq:ccqq}) with 
$n={}^1\!S_0^{(8)},{}^3\!S_1^{(8)},{}^3\!P_J^{(8)}$, where
they are always related to $\gamma\to q\overline{q}$ splitting.
The corresponding expressions for the Eikonal factors $F_{ij}$, where $i=1,2$
and $j=3,4$, may be found in Table~\ref{tab:f}.
They are related to the spacelike $\gamma\to q$ splitting function,
\begin{equation}
P_{\gamma\to q}(z)=\frac{e_q^2N_c}{T_F}P_{g\to q}(z),
\end{equation}
where $P_{g\to q}(z)$ is defined in Eq.~(\ref{eq:pgq}), as
\begin{equation}
F_{ij}=-\frac{2e^2}{zs_{ij}}\left[\frac{P_{\gamma\to q}(z)}{N_c}-\epsilon
e_q^2 \right].
\end{equation}

The integration of Eq.~(\ref{eq:ini}) over the collinear region of the phase
space, with $s_{23}<\delta_i s$, can be performed analytically.
In terms of differential cross sections, the result reads
\begin{eqnarray}
d\sigma_c^{23}&=&\int_0^1 dz 
\frac{\alpha}{2\pi}\left(-\frac{1}{\epsilon}\right)
\frac{\Gamma(1-\epsilon)}{\Gamma(1-2\epsilon)}
\left(\frac{4\pi\mu^2 }{(1-z)\delta_is}\right)^\epsilon
[P_{\gamma\to q}(z) - \epsilon e_q^2 N_c]\nonumber\\
&&{}\times d\sigma(\gamma(k_1)+\overline{q}(zk_2)\to
c\overline{c}[n](p)+\overline{q}(k_4)),
\label{eq:int}
\end{eqnarray}
where $s$ is the two-photon invariant mass square defined above 
Eq.~(\ref{eq:stu}).
According to the mass factorization theorem \cite{dew}, the form of this
collinear singularity connected with an incoming-photon leg is universal and
can be absorbed into the bare PDF of the antiquark $\overline{q}$ inside the
resolved photon, $f_{\overline{q}/\gamma}(z)$.
To this end, one defines the renormalized photon PDF as
\begin{equation}
f_{\overline{q}/\gamma}(z,M^2)=f_{\overline{q}/\gamma}(z)
+\frac{\alpha}{2\pi}\left[-\frac{1}{\epsilon}\,
\frac{\Gamma(1-\epsilon)}{\Gamma(1-2\epsilon)}
\left(\frac{4\pi\mu^2}{M^2}\right)^\epsilon
P_{\gamma\to q}(z)+e_q^2C_\gamma(z)\right],
\label{eq:pdf}
\end{equation}
where $M$ is the factorization scale and the finite function $C_\gamma(z)$ is
introduced to fix the factorization scheme.
In the modified minimal-subtraction ($\overline{\rm MS}$) scheme, we have
$C_\gamma(z)=0$, while in the so-called DIS${}_\gamma$ scheme, which is used,
e.g., in Ref.~\cite{grs}, we have
\begin{equation}
C_\gamma(z)
=N_c\left\{\left[z^2+(1-z)^2\right]\ln\frac{1-z}{z}+8z(1-z)-1\right\}.
\end{equation}
Incorporating the second term on the right-hand side of Eq.~(\ref{eq:pdf}) in
Eq.~(\ref{eq:int}), we are left with the finite contribution,
\begin{eqnarray}
d\sigma_c^{23}&=&\int_0^1 dz \frac{\alpha}{2\pi}\left\{
P_{\gamma\to\overline{q}}(z)\ln\frac{(1-z)\delta_is}{ M^2}
+e_q^2[N_c-C_\gamma(z)]\right\}\nonumber\\
&&{}\times d\sigma(\gamma(k_1)+\overline{q}(zk_2)\to
c\overline{c}[n](p)+\overline{q}(k_4)),
\end{eqnarray}
which depends on the choice of $M$. 
In turn, the resolved-photon contribution is evaluated with the renormalized
photon PDF $f_{\overline{q}/\gamma}(z,M^2)$, which also depends on $M$.
In the sum of these two contributions, this $M$ dependence cancels up to terms
beyond NLO.
In this sense, the notion of direct-photon contribution ceases to be 
separately well defined at NLO.

\subsection{Assembly of the NLO cross section}

The NLO result for the cross section of process~(\ref{eq:ccg}) is obtained by
adding the virtual and real corrections to the LO result.
Collecting the various contributions discussed above, arising from the
parameter and wave-function renormalization (ct), the operator redefinition
(op), the initial-state (is) and final-state (fs) collinear configurations,
the soft-gluon radiation (so), and the hard-parton emission (ha), we can
schematically write the resulting differential cross section of
$\gamma\gamma\to H+X$, including the MEs, as
\begin{eqnarray}
d\sigma(\mu,\lambda,M)&=&
d\sigma_0(\mu,\lambda)[1
+\delta_{\rm vi}(\mu;\epsilon_{\rm UV},\epsilon_{\rm IR},v)
+\delta_{\rm ct}(\mu;\epsilon_{\rm UV},\epsilon_{\rm IR})
+\delta_{\rm op}(\mu,\lambda;\epsilon_{\rm IR},v)\nonumber\\
&&{}+\delta_{\rm fs}(\mu;\epsilon_{\rm IR},\delta_f)]
+d\sigma_{\rm is}(\mu,\lambda,M;\delta_i)
+d\sigma_{\rm so}(\mu,\lambda;\epsilon_{\rm IR},\delta_f)\nonumber\\
&&{}+d\sigma_{\rm ha}(\mu,\lambda;\delta_i,\delta_f),
\label{eq:sum}
\end{eqnarray}
where the dependences on the unphysical mass scales $\mu$, $\lambda$, and $M$
and the regulators $\epsilon_{\rm IR}$, $\epsilon_{\rm UV}$, $v$, $\delta_i$,
and $\delta_f$ are indicated in parentheses for each term.
Everywhere in Eq.~(\ref{eq:sum}), $\alpha_s(\mu)$ is evaluated in the
$\overline{\rm MS}$ scheme using the two-loop formula~(\ref{eq:as}), $m$ is
defined in the OS scheme, and the MEs are understood as their
$\overline{\rm MS}$ values $\langle{\cal O}^H[n]\rangle_r(\lambda)$.
This is necessary to ensure the exact cancellation of the singularities.

The right-hand side of Eq.~(\ref{eq:sum}) is manifestly finite.
The UV divergences cancel between $\delta_{\rm vi}$ and $\delta_{\rm ct}$;
the IR singularities among $\delta_{\rm vi}$, $\delta_{\rm ct}$,
$\delta_{\rm op}$, $\delta_{\rm fs}$, and $d\sigma_{\rm so}$; and the Coulomb
singularities between $\delta_{\rm vi}$ and $\delta_{\rm op}$.
Note that $\delta_{\rm op}$ is UV finite upon operator renormalization and
that $d\sigma_{\rm is}$ is IR finite upon photon PDF renormalization;
therefore, $\delta_{\rm op}$ and $d\sigma_{\rm is}$ in Eq.~(\ref{eq:sum}) do
not depend any more on $\epsilon_{\rm UV}$ and $\epsilon_{\rm IR}$,
respectively.
The full details about the cancellation of the IR singularities in the various
$c\overline{c}$ Fock states $n$ are presented in Table~\ref{tab:ir}, where the
following short-hand notation is used:
\begin{eqnarray}
T_{\gamma q\to c\overline{c}q}[n]&=&
\frac{\alpha}{2\pi}\,
\frac{\Gamma(1-\epsilon)}{\Gamma(1-2\epsilon)} 
\overline{|{\cal T}_0|^2}(\gamma q\to c\overline{c}[n]+q).
\end{eqnarray}

\begin{table}[ht]
\begin{center}
\begin{tabular}{|l|l|l|}
\hline\hline
Subprocess & Source & IR-singular term \\
\hline
$\gamma\gamma\to c\bar{c}\left[{}^3\!S_1^{(8)}\right]+g$
& virtual &
$\frac{\alpha_s}{\pi}
\left(\frac{4\pi\mu^2}{m^2}\right)^\epsilon\exp(-\epsilon\gamma_E)
\left[-\frac{C_A}{\epsilon}\left(\frac{1}{2\epsilon}
-\ln\frac{s_1}{2m^2}+\frac{17}{12}\right)\right.$
\\
&&
${}+\left.\frac{T_Fn_f}{3\epsilon}
+\left(C_F-\frac{C_A}{2}\right)\frac{\pi^2}{2v}\right] 
\overline{|{\cal T}_0|^2}$
\\
& operator &
$-\left(C_F-\frac{C_A}{2}\right)\frac{\pi\alpha_s}{2v}
\left\langle{\cal O}\left[{}^3\!S_1^{(8)}\right]\right\rangle$
\\
$\gamma\gamma\to c\overline{c}\left[{}^3\!P_J^{(1)}\right]+gg$
& operator &
$\frac{4C_F\alpha_s}{3\pi m^2}
\left(\frac{4\pi\mu^2}{\lambda^2}\right)^\epsilon\exp(-\epsilon\gamma_E)
\frac{1}{\epsilon}
\left\langle{\cal O}\left[{}^3\!P_J^{(1)}\right]\right\rangle$
\\
& soft &
$-\frac{4C_F\alpha_s}{3\pi m^2}
\left(\frac{4\pi\mu^2}{m^2}\right)^\epsilon\exp(-\epsilon\gamma_E)
\frac{1}{\epsilon}\overline{|{\cal T}_0|^2}$
\\
$\gamma\gamma\to c\bar{c}\left[{}^3\!S_1^{(8)}\right]+gg$
& final-state &
$\frac{C_A\alpha_s}{\pi}
\left(\frac{4\pi\mu^2}{m^2}\right)^\epsilon\exp(-\epsilon\gamma_E)
\frac{1}{\epsilon}\left(
\ln\frac{\delta_f s}{s_1}+\frac{11}{12}\right)\overline{|{\cal T}_0|^2}$
\\
& soft &
$\frac{C_A\alpha_s}{\pi}
\left(\frac{4\pi\mu^2}{m^2}\right)^\epsilon\exp(-\epsilon\gamma_E)
\frac{1}{\epsilon}\left(\frac{1}{2\epsilon}
-\ln\frac{\delta_f s}{2m^2}+\frac{1}{2}\right)\overline{|{\cal T}_0|^2}$
\\
$\gamma\gamma\to c\bar{c}[{}^3\!P_J^{(8)}]+gg$
& operator &
$\frac{4B_F\alpha_s}{3\pi m^2}
\left(\frac{4\pi\mu^2}{\lambda^2}\right)^\epsilon\exp(-\epsilon\gamma_E)
\frac{1}{\epsilon}
\left\langle{\cal O}\left[{}^3\!P_J^{(8)}\right]\right\rangle$
\\
& soft &
$-\frac{4B_F\alpha_s}{3\pi m^2}
\left(\frac{4\pi\mu^2}{m^2}\right)^\epsilon\exp(-\epsilon\gamma_E)
\frac{1}{\epsilon}\overline{|{\cal T}_0|^2}$
\\ 
$\gamma\gamma\to c\bar{c}\left[{}^3\!S_1^{(8)}\right]+q\bar{q}$
& final-state &
$-\frac{T_Fn_f\alpha_s}{3\pi}
\left(\frac{4\pi\mu^2}{m^2}\right)^\epsilon\exp(-\epsilon\gamma_E)
\frac{1}{\epsilon}\overline{|{\cal T}_0|^2}$
\\
$\gamma\gamma\to c\bar{c}[n]+q\bar{q}$
& initial-state & 
$-\left(\frac{4\pi \mu^2}{(1-z)\delta_is}\right)^\epsilon  
\frac{1}{\epsilon}P_{\gamma\to q}(z)
T_{\gamma q\to c\overline{c}q}[n]$
\\
$\gamma q\to c\bar{c}[n]+q$
& mass fact. & 
$\left(\frac{4\pi \mu^2}{M^2}\right)^\epsilon  
\frac{1}{\epsilon}P_{\gamma\to q}(z)
T_{\gamma q\to c\overline{c}q}[n]$
\\
\hline\hline
\end{tabular}
\caption{Compilation of the IR-singular terms arising from the various sources
in the various partonic subprocesses.
Here, $n={}^1\!S_0^{(8)},{}^3\!S_1^{(8)},{}^3\!P_J^{(8)}$.}
\label{tab:ir}
\end{center}
\end{table}

The right-hand side of Eq.~(\ref{eq:sum}) is independent of the cut-off
parameters $\delta_i$ and $\delta_f$.
Specifically, the $\delta_i$ dependence cancels between $d\sigma_{\rm is}$ and
$d\sigma_{\rm ha}$ and the $\delta_f$ dependence among $\delta_{\rm fs}$,
$d\sigma_{\rm so}$, and $d\sigma_{\rm ha}$.
While $\delta_{\rm fs}$, $d\sigma_{\rm is}$, and $d\sigma_{\rm so}$ are known
in analytic form, the phase-space integrals occurring in the evaluation of
$d\sigma_{\rm ha}$ are rather cumbersome and are thus solved numerically.
Consequently, the cancellation of $\delta_i$ and $\delta_f$ has to be
established numerically, too.
The goodness of this cancellation is assessed in Section~\ref{sec:num}.

The right-hand side of Eq.~(\ref{eq:sum}) is independent of the
renormalization scales $\mu$ and $\lambda$ up to terms that are formally
beyond NLO;
this cancellation is not exact because the running of $\alpha_s(\mu)$ and
$\langle{\cal O}^H[n]\rangle_r(\lambda)$ is determined from the respective
renormalization group equations, which resum logarithmic corrections to all
orders.
However, the right-hand side of Eq.~(\ref{eq:sum}) depends on the
factorization scale $M$ already in NLO;
this $M$ dependence is cancelled up to terms beyond NLO once the cross section
of single-resolved photoproduction, which depends on $M$ through the photon
PDFs $f_{a/\gamma}(x_a,M)$, is added.
If $M$ is to be varied, this must thus be done simultaneously in the cross
sections of direct and single-resolved photoproduction.

\section{Numerical results}
\label{sec:num}

We are now in a position to present our numerical analysis of the inclusive
production of prompt $J/\psi$ mesons in two-photon collisions with direct
photon interactions at NLO in the NRQCD factorization framework.
We consider TESLA in its $e^+e^-$ mode with $\sqrt s=500$~GeV, where the
photons are produced via bremsstrahlung and beamstrahlung.

The discussion of the numerical analysis proceeds in three steps.
We first specify our input parameters.
We then verify that our numerical evaluation is independent of the technical
cut-off parameters $\delta_i$ and $\delta_f$ and assess its dependence on
the renormalization and factorization scales $\mu$, $\lambda$, and $M$.
Finally, we explore the phenomenological consequences of our analysis by
studying the size and impact of the NLO corrections on the $p_T$ and $y$
distributions of the cross section.

\subsection{Input parameters}

We use $m=1.5$~GeV and $\alpha=1/137.036$.
For direct photoproduction at NLO (LO), we employ the two-loop (one-loop)
formula for $\alpha_s^{(n_f)}(\mu)$ \cite{pdg} with $n_f=3$ active quark
flavours and $\Lambda_{\rm QCD}^{(3)}=299$~MeV (204~MeV) \cite{grs}.
We use the photon PDFs from Gl\"uck, Reya, and Schienbein (GRS) \cite{grs},
which are the only available ones that are implemented in the
fixed-flavour-number scheme, with $n_f=3$.
When we combine the contribution due to single-resolved photoproduction, which
is so far only known at LO, with the NLO one due to direct photoproduction,
then we nevertheless evaluate it using the NLO formula for
$\alpha_s^{(n_f)}(\mu)$ and the NLO set of the GRS photon PDFs so as to
maximize the compensation of the $M$ dependence.
In order to render this choice fully consistent, we will have to include the
NLO correction to the single-resolved contribution and also the double-resolved
contribution at NLO, once they become available.
On the other hand, we consistently evaluate the single-resolved contribution
using the LO formula for $\alpha_s^{(n_f)}(\mu)$ and the LO set of the GRS
photon PDFs when we consider it separately or in combination with the direct
contribution at LO.
Our default choice of renormalization and factorization scales is $\mu=M=m_T$
and $\lambda=m$.
In want of NLO sets of $J/\psi$, $\chi_{cJ}$, and $\psi^\prime$ MEs, we adopt
the LO sets determined in Ref.~\cite{bkl} using the LO set of proton PDFs from
Martin, Roberts, Stirling, and Thorne (MRST98LO) \cite{mrst}.
Specifically,
$\left\langle{\cal O}^{\psi(nS)}\left[{}^3\!S_1^{(1)}\right]\right\rangle$ and
$\left\langle{\cal O}^{\chi_{c0}}\left[{}^3\!P_0^{(1)}\right]\right\rangle$
were extracted from the measured partial decay widths of $\psi(nS)\to l^+l^-$
and $\chi_{c2}\to\gamma\gamma$ \cite{pdg}, respectively, while
$\left\langle{\cal O}^{\psi(nS)}\left[{}^1\!S_0^{(8)}\right]\right\rangle$,
$\left\langle{\cal O}^{\psi(nS)}\left[{}^3\!S_1^{(8)}\right]\right\rangle$,
$\left\langle{\cal O}^{\psi(nS)}\left[{}^3\!P_0^{(8)}\right]\right\rangle$,
and
$\left\langle{\cal O}^{\chi_{c0}}\left[{}^3\!S_1^{(8)}\right]\right\rangle$
were fitted to the transverse-momentum distributions of $\psi(nS)$ and
$\chi_{cJ}$ inclusive hadroproduction \cite{abe} and the cross-section ratio
$\sigma_{\chi_{c2}}/\sigma_{\chi_{c1}}$ \cite{aff} measured at the Tevatron.
The fit results for
$\left\langle{\cal O}^{\psi(nS)}\left[{}^1\!S_0^{(8)}\right]\right\rangle$ and
$\left\langle{\cal O}^{\psi(nS)}\left[{}^3\!P_0^{(8)}\right]\right\rangle$ are
strongly correlated, so that the linear combination
\begin{equation}
M_r^{\psi(nS)}
=\left\langle{\cal O}^{\psi(nS)}\left[{}^1\!S_0^{(8)}\right]\right\rangle
+\frac{r}{m^2}
\left\langle{\cal O}^{\psi(nS)}\left[{}^3\!P_0^{(8)}\right]\right\rangle,
\label{eq:mr}
\end{equation}
with a suitable value of $r$, is quoted.
Unfortunately, Eq.~(\ref{eq:sum}) is sensitive to linear combination of
$\left\langle{\cal O}^{\psi(nS)}\left[{}^1\!S_0^{(8)}\right]\right\rangle$ and
$\left\langle{\cal O}^{\psi(nS)}\left[{}^3\!P_0^{(8)}\right]\right\rangle$
that is different from the one appearing in Eq.~(\ref{eq:mr}).
In want of more specific information, we thus make the democratic choice
$\left\langle{\cal O}^{\psi(nS)}\left[{}^1\!S_0^{(8)}\right]\right\rangle
=\left(r/m^2\right)
\left\langle{\cal O}^{\psi(nS)}\left[{}^3\!P_0^{(8)}\right]\right\rangle
=M_r^{\psi(nS)}/2$.

We now discuss the photon flux functions that enter our predictions for
photoproduction in the $e^+e^-$ mode of TESLA.
The energy spectrum of the bremsstrahlung photons is well described in the WWA
by Eq.~(27) of Ref.~\cite{fri}.
We assume that the scattered electrons and positrons will be antitagged, as
was usually the case at LEP2, and take the maximum scattering angle to be
$\theta_{\rm max}=25$~mrad \cite{theta}.
The energy spectrum of the beamstrahlung photons is approximately described by
Eq.~(2.14) of Ref.~\cite{pes}.
It is controlled by the effective beamstrahlung parameter $\Upsilon$, which is
given by Eq.~(2.10) of that reference.
Inserting the relevant TESLA parameters for the $\sqrt S=500$~GeV baseline
design specified in Table~1.3.1 of Ref.~\cite{tesla} in that formula, we
obtain $\Upsilon=0.053$.
We coherently superimpose the WWA and beamstrahlung spectra.

With the NLO corrections to single- and double-resolved photoproduction yet to
be evaluated, we are not in a position to present a complete phenomenological
prediction that could be confronted with experimental data as it stands.
Therefore, we refrain from presenting a full-fledged quantitative estimate of
the theoretical uncertainties.
However, in the next section, we do investigate the dependences on the
renormalization and factorization scales $\mu$, $\lambda$, and $M$.

\subsection{Academic study}

For the purpose of the following technical study, it is sufficient to consider
a typical kinematic situation.
We thus choose as our reference quantity the differential cross section
$d^2\sigma/dp_T\,dy$ at $p_T=5$~GeV and $y=0$.

\begin{figure}[ht]
\begin{center}
\epsfig{figure=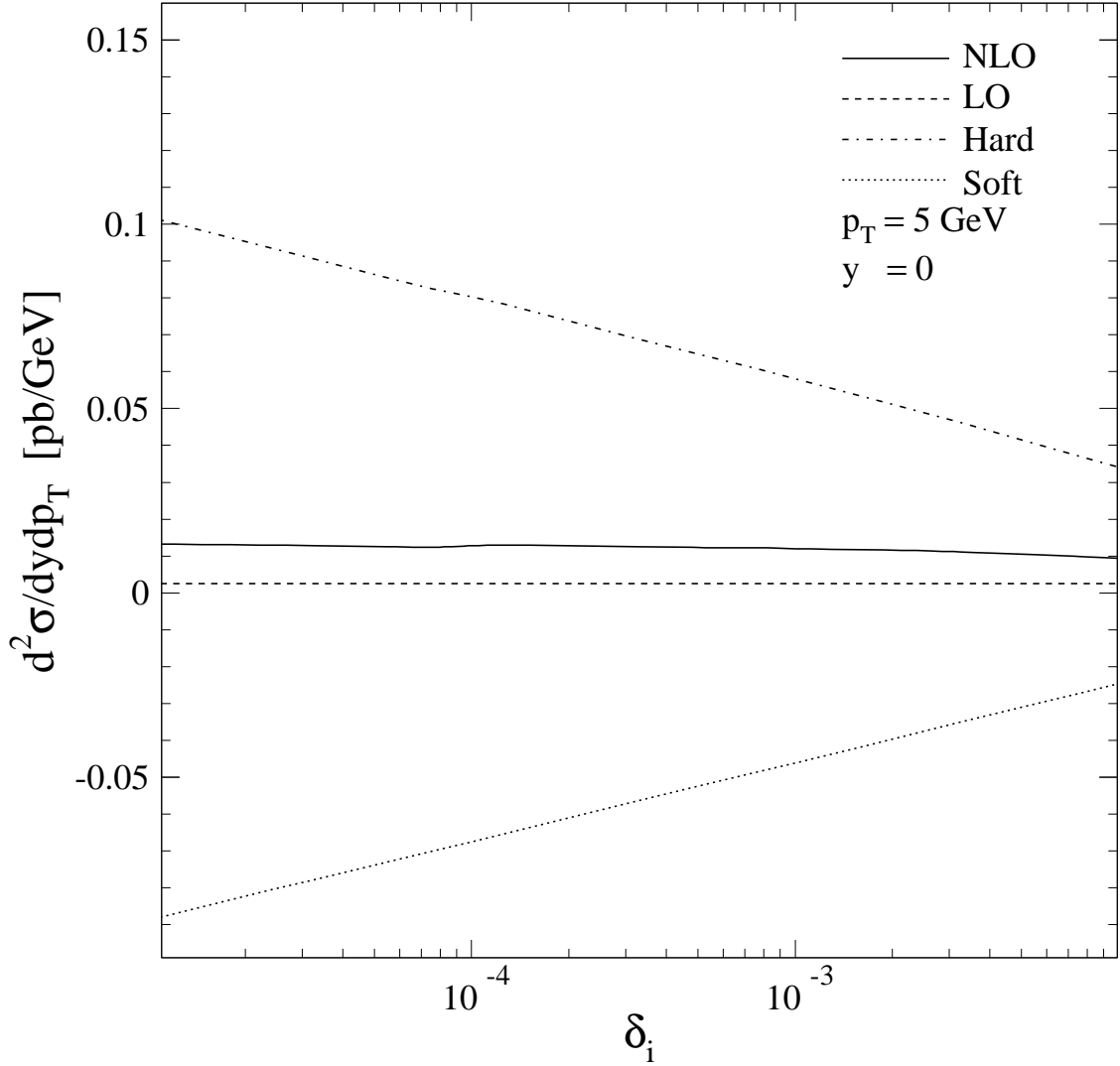,width=\textwidth}
(a)\\
\caption{Differential cross section $d^2\sigma/dp_T\,dy$ in pb/GeV of
$e^+e^-\to e^+e^-J/\psi+X$ in direct photoproduction at TESLA with
$\sqrt s=500$~GeV for prompt $J/\psi$ mesons with $p_T=5$~GeV and $y=0$.
The NLO result (solid lines), its hard, noncollinear component (dot-dashed
lines), which corresponds to $d\sigma_{\rm ha}$ in Eq.~(\ref{eq:sum}), and the
remainder (dotted lines) are shown (a) as functions of $\delta_i$ for
$\delta_f=5\times10^{-3}$ and (b) as functions of $\delta_f$ for
$\delta_i=10^{-4}$.
For comparison, also the LO result (dashed lines) is shown.}
\label{fig:cut}
\end{center}
\end{figure}

In Figs.~\ref{fig:cut}(a) and (b), the NLO result of direct photoproduction
(solid lines), its hard, noncollinear component (dot-dashed lines), which
corresponds to $d\sigma_{\rm ha}$ in Eq.~(\ref{eq:sum}), and the remainder
(dotted lines) are plotted as functions of $\delta_i$ for
$\delta_f=5\times10^{-3}$ and as functions of $\delta_f$ for
$\delta_i=5\times10^{-4}$, respectively.
For comparison, also the LO result (dashed lines), which is, of course,
independent of $\delta_i$ and $\delta_f$, is shown.
Notice that $\delta_i$ and $\delta_f$ are varied over several orders of
magnitude.
They have to be chosen judiciously: if they are chosen too large, then the
collinear approximation underlying $\delta_{\rm fs}$ and $d\sigma_{\rm is}$
and the soft approximation underlying $d\sigma_{\rm so}$ break down; if they
are chosen too small, then the numerical evaluation of $d\sigma_{\rm ha}$
becomes inaccurate.
Our default values for the rest of the numerical analysis are
$\delta_i=5\times10^{-4}$ and $ \delta_f=5\times10^{-3}$.
Since the remainder depends logarithmically on $\delta_i$ and $\delta_f$, it
is represented by straight lines in Figs.~\ref{fig:cut}(a) and (b), the
abscissae of which are logarithmic.
The same is true, to good approximation, for the hard, noncollinear component,
and the combined NLO result is practically independent of $\delta_i$ and
$\delta_f$, as it should.
This suggests that the numerical integration is rather precise and stable.
We conclude that this error source can be safely neglected in the
determination of the theoretical uncertainty of the NLO result.

\begin{figure}[ht]
\begin{center}
\epsfig{figure=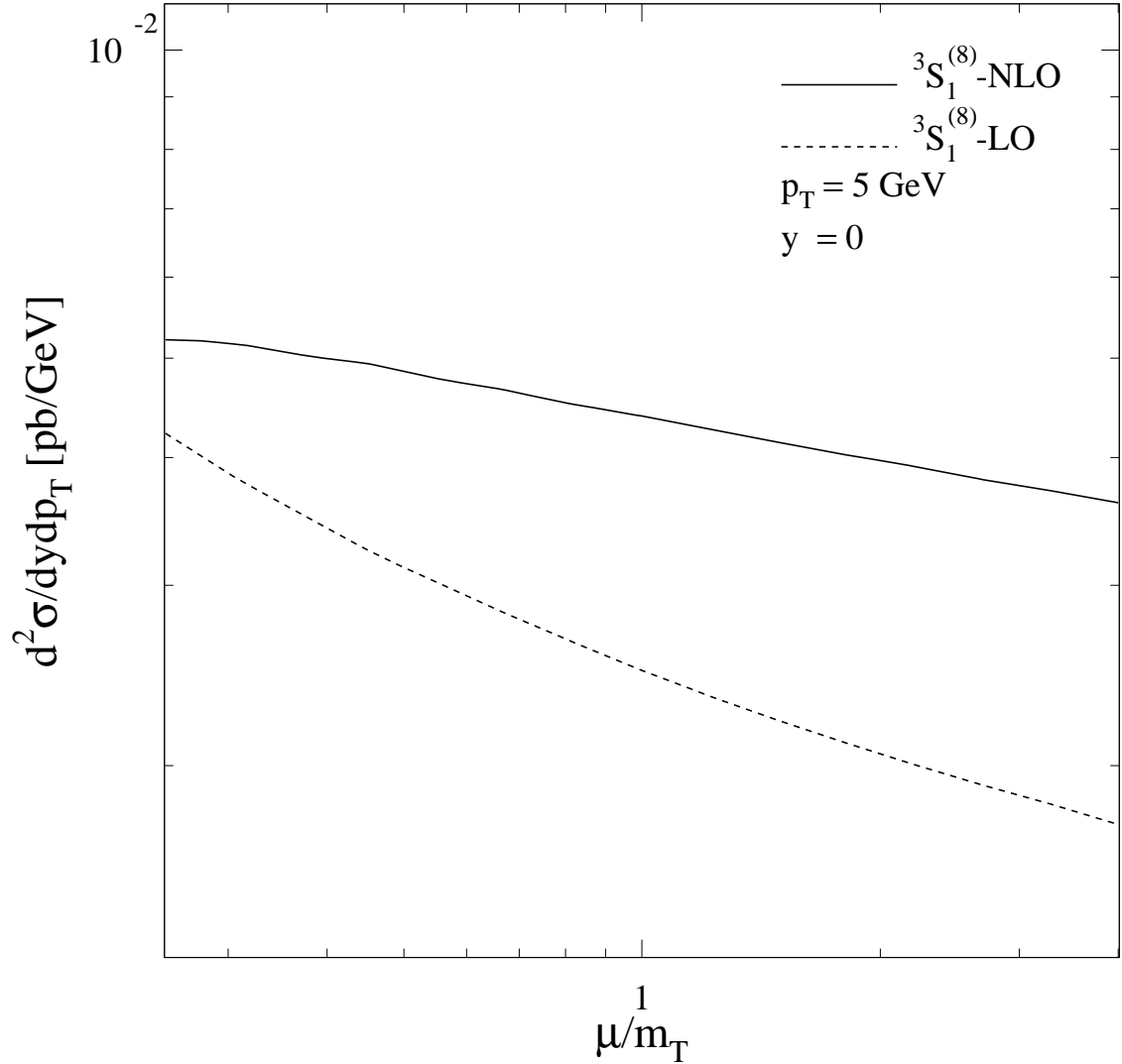,width=\textwidth}
\caption{Differential cross section $d^2\sigma/dp_T\,dy$ in pb/GeV of
$e^+e^-\to e^+e^-J/\psi+X$ in direct photoproduction at TESLA with
$\sqrt s=500$~GeV for prompt $J/\psi$ mesons with $p_T=5$~GeV and $y=0$.
The NLO (solid line) and LO (dashed line) contributions due to the
$c\overline{c}$ Fock state $n={}^3\!S_1^{(8)}$ are shown as functions of
$\mu$.}
\label{fig:mu}
\end{center}
\end{figure}

In Fig.~\ref{fig:mu}, the NLO (solid line) and LO (dashed line) contributions
due to the $c\overline{c}$ Fock state $n={}^3\!S_1^{(8)}$ are shown as
functions of $\mu$, while $\lambda$ and $M$ are kept fixed at their reference
values.
We focus on this particular channel because it is the only one already open at
LO, so that a compensation of the $\mu$ dependence can occur.
Notice that $\mu$ is varied by more than one order of magnitude, from $m_T/4$
to $4m_T$.
Passing from LO to NLO, the $\mu$ dependence is appreciably reduced, 
reflecting
the partial compensation of the $\mu$ dependence of $d\sigma_0$ by the one of
$\delta_{\rm ct}$ in Eq.~(\ref{eq:sum}). 
In fact, the related theoretical uncertainty amounts to ${+70\atop-30}\%$ at
LO and to $\pm 18\%$ at NLO.

\begin{figure}[ht]
\begin{center}
\epsfig{figure=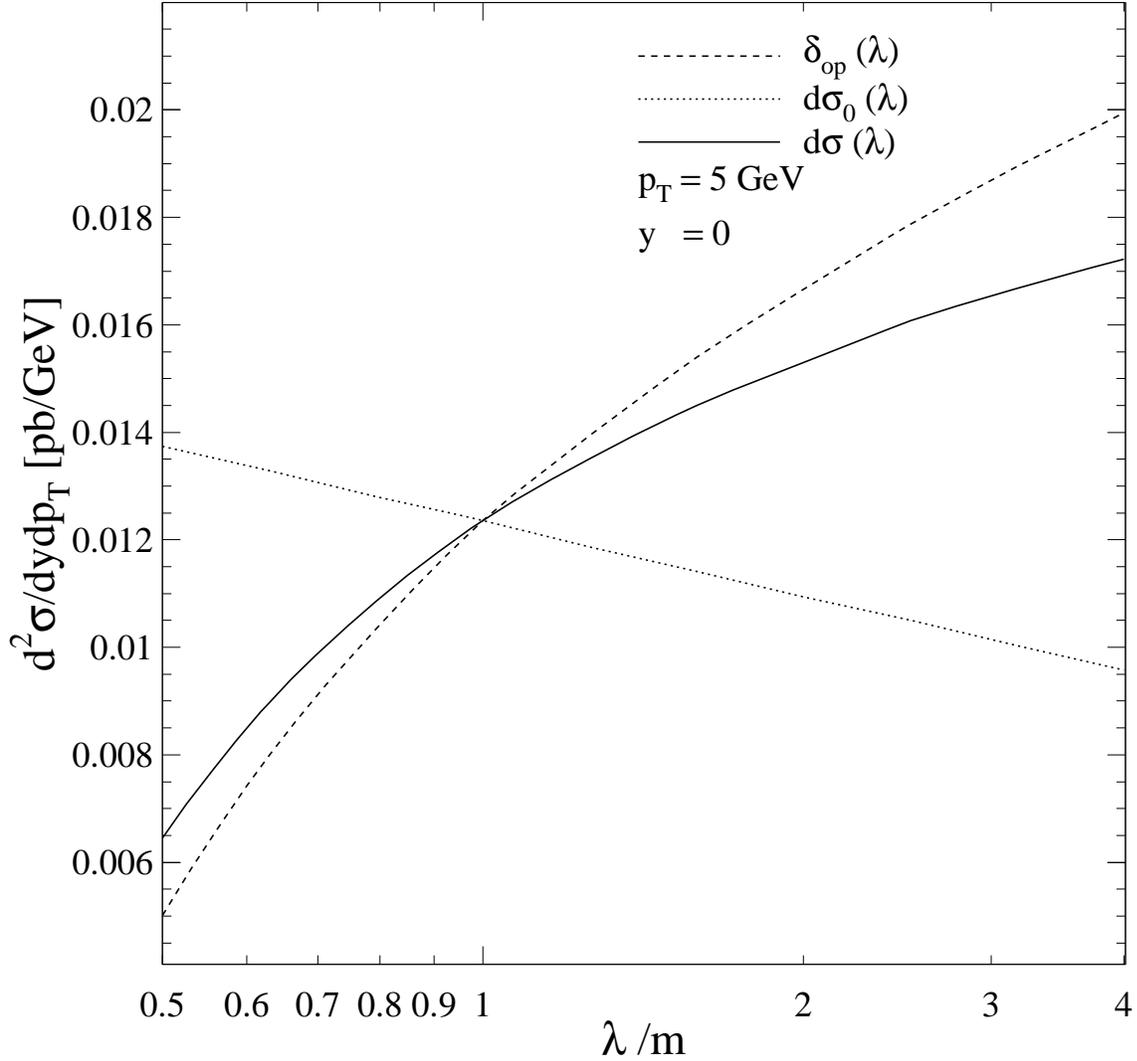,width=\textwidth}
\caption{Differential cross section $d^2\sigma/dp_T\,dy$ in pb/GeV of
$e^+e^-\to e^+e^-J/\psi+X$ in direct photoproduction at TESLA with
$\sqrt s=500$~GeV for prompt $J/\psi$ mesons with $p_T=5$~GeV and $y=0$.
The NLO result is shown as a function of $\lambda$,
(i) keeping $\lambda$ in the partonic cross sections fixed (dashed line),
(ii) keeping $\lambda$ in
$\left\langle{\cal O}^H\left[{}^3\!S_1^{(8)}\right]\right\rangle_r(\lambda)$
fixed (dotted line), and (iii) varying all occurrences of $\lambda$
simultaneously (solid line).}
\label{fig:la}
\end{center}
\end{figure}

In Fig.~\ref{fig:la}, the NLO result is shown as a function of $\lambda$
(solid line), while $\mu$ and $M$ are kept fixed at their reference values.
Notice that $\lambda$ is varied by almost one order of magnitude, from $m/2$
to $4m$.
The related theoretical uncertainty amounts to $\pm 50\%$.
The $\lambda$-dependent terms in  Eq.~(\ref{eq:sum}) are $d\sigma_0$ and
$\delta_{\rm op}$.
In order to exhibit the partial compensation in $\lambda$ dependence between
these two terms, we also include in Fig.~\ref{fig:la} the results that are
obtained by only varying $\lambda$ in $d\sigma_0$ (dashed line) or
$\delta_{\rm op}$ (dotted line) at a time.
The dotted line is straight, reflecting the fact that $\delta_{\rm op}$
depends on $\lambda$ through a single logarithm.

\begin{figure}[ht]
\begin{center}
\epsfig{figure=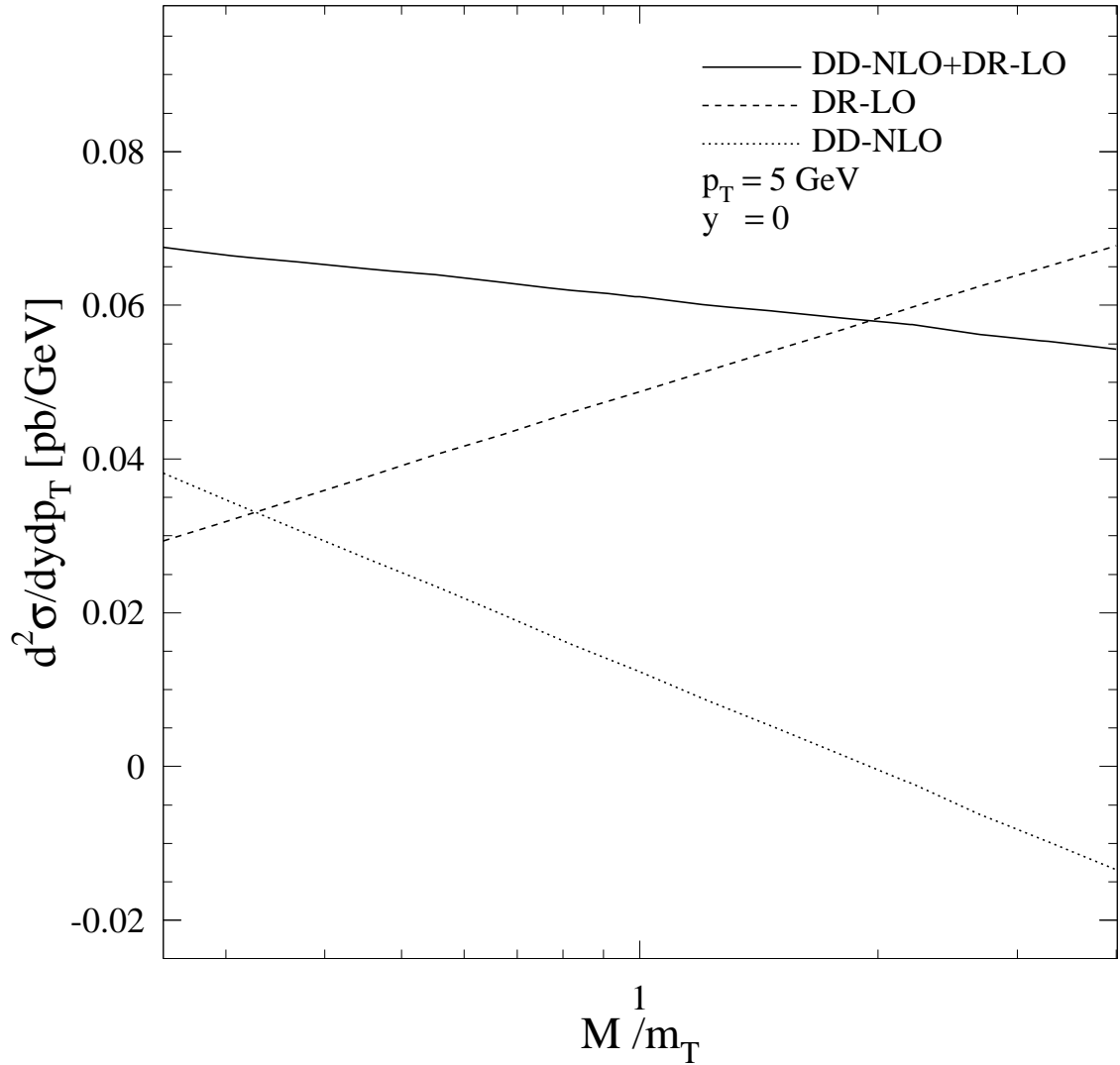,width=\textwidth}
\caption{Differential cross section $d^2\sigma/dp_T\,dy$ in pb/GeV of
$e^+e^-\to e^+e^-J/\psi+X$ at TESLA with $\sqrt s=500$~GeV for prompt $J/\psi$
mesons with $p_T=5$~GeV and $y=0$.
The NLO result of direct photoproduction (dotted line), the LO result of
single-resolved photoproduction (dashed line), and their sum (solid line) are
shown as functions of $M$.}
\label{fig:m}
\end{center}
\end{figure}

In Fig.~\ref{fig:m}, the NLO result of direct photoproduction (dotted line),
the LO result of single-resolved photoproduction evaluated with the NLO
versions of $\alpha_s^{(n_f)}(\mu)$ and the photon PDFs (dashed line), and
their sum (solid line) are shown as functions of $M$, while $\mu$ and
$\lambda$ are kept fixed at their reference values.
Notice that $M$ is varied by more than one order of magnitude, from $m_T/4$
to $4m_T$.
The $M$ dependences of the individual contributions almost cancel each other,
leaving a theoretical uncertainty of $\pm 11\%$ on their sum.

\subsection{Phenomenological study}

\begin{figure}[ht]
\begin{center}
\epsfig{figure=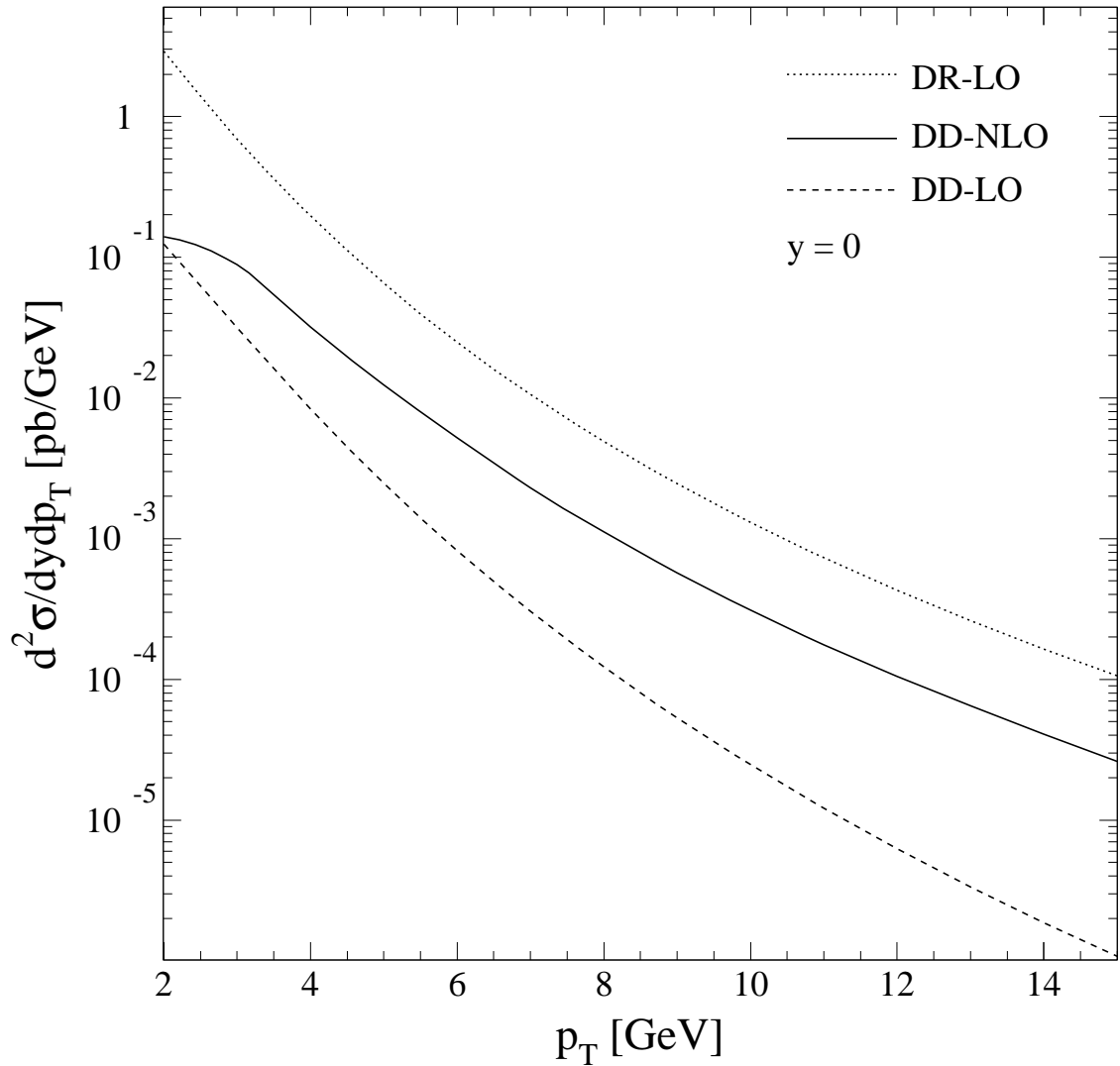,width=\textwidth}
(a)\\
\caption{Differential cross section $d^2\sigma/dp_T\,dy$ in pb/GeV of
$e^+e^-\to e^+e^-J/\psi+X$ at TESLA with $\sqrt s=500$~GeV for prompt $J/\psi$
mesons (a) with $y=0$ as a function of $p_T$ and (b) with $p_T=5$~GeV as a
function $y$.
The LO (dashed line) and NLO (solid line) results of direct photoproduction
are compared with the LO result of single-resolved photoproduction (dotted
line).}
\label{fig:xs}
\end{center}
\end{figure}

In Fig.~\ref{fig:xs}, we study $d^2\sigma/dp_T\,dy$ (a) for $y=0$ as a
function of $p_T$ and (b) for $p_T=5$~GeV as a function of $y$.
In each case, the LO (dashed line) and NLO (solid line) results of direct
photoproduction as well as the LO result of single-resolved photoproduction
evaluated with the LO versions of $\alpha_s^{(n_f)}(\mu)$ and the photon PDFs
(dotted line) are shown.
Notice that our analysis is only valid for finite values of $p_T$; in the
limit $p_T\to0$, additional IR singularities occur, which require a more
sophisticated scheme of phase space slicing.
Therefore, we do not consider $p_T$ values below 2~GeV in 
Fig.~\ref{fig:xs}(a).
For a more detailed discussion of this point, we refer to Ref.~\cite{gg}.

\begin{figure}[ht]
\begin{center}
\epsfig{figure=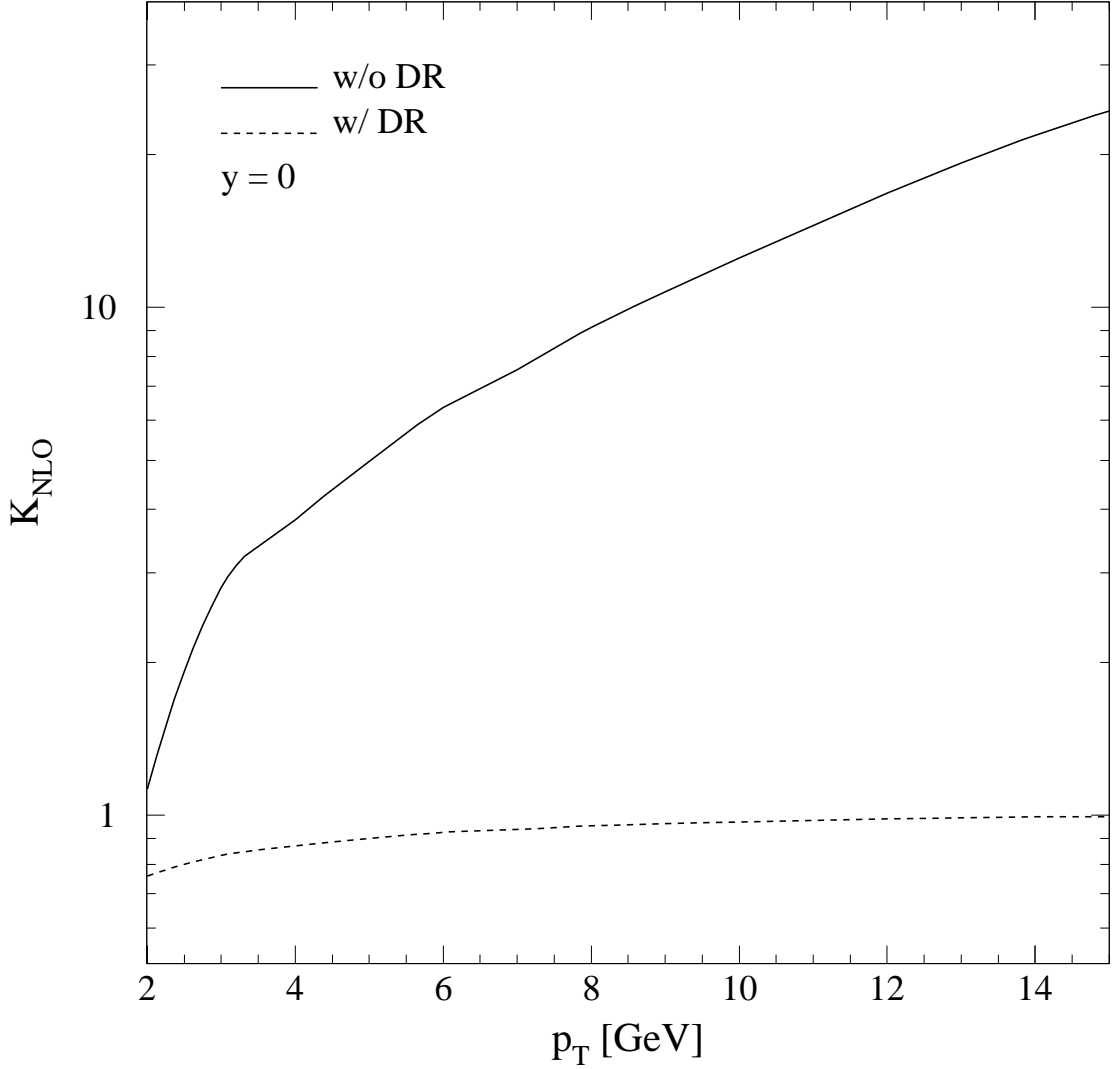,width=\textwidth}
(a)\\
\caption{QCD correction factor $K$ of the differential cross section
$d^2\sigma/dp_T\,dy$ of $e^+e^-\to e^+e^-J/\psi+X$ at TESLA with
$\sqrt s=500$~GeV for prompt $J/\psi$ mesons (a) with $y=0$ as a function of
$p_T$ and (b) with $p_T=5$~GeV as a function $y$.
$K$ is defined as (i) the ratio of the NLO and LO results of direct
photoproduction (solid lines) and (ii) the same quantity after adding the LO
result of single-resolved photoproduction to both numerator and denominator,
where it is evaluated with the NLO and LO versions of $\alpha_s^{(n_f)}(\mu)$
and the photon PDFs, respectively (dashed lines).}
\label{fig:k}
\end{center}
\end{figure}

From Fig.~\ref{fig:xs}(a), we observe that, with increasing value of $p_T$,
the NLO result of direct photoproduction falls of considerably more slowly
than the LO one.
This is also evident from Fig.~\ref{fig:k}(a), where the NLO to LO ratio is
shown for $y=0$ as a function of $p_T$ (solid line).
This feature may be understood by observing that so-called
{\it fragmentation-prone} partonic subprocesses \cite{jb} start to contribute
to direct photoproduction at NLO, while they are absent at LO.
Such subprocesses contain a gluon with small virtuality, $q^2=4m^2$, that
splits into a $c\overline{c}$ pair in the Fock state $n={}^3\!S_1^{(8)}$ and
thus generally generate dominant contributions at $p_T\gg2m$ due to the
presence of a large gluon propagator.
In the case under consideration, the relevant $T$-matrix elements are those of
$\gamma\gamma\to c\overline{c}\left[{}^3\!S_1^{(8)}\right]g$ and
$\gamma\gamma\to c\overline{c}\left[{}^3\!S_1^{(8)}\right]q\overline{q}$ that
are proportional to $e_q^2$;
the corresponding diagrams are the last two ones in Fig.~\ref{fig:loop} and
the last six ones in Fig.~\ref{fig:ccqq}, respectively.
In single-resolved photoproduction, a fragmentation-prone partonic subprocess
already contributes at LO;
the relevant $T$-matrix elements is the one of
$\gamma q\to c\overline{c}\left[{}^3\!S_1^{(8)}\right]q$ that is proportional
to $e_q$.
This explains why the solid and dotted curves in Fig.~\ref{fig:xs}(a) run
parallel in the upper $p_T$ range.
At low values of $p_T$, the fragmentation-prone partonic subprocesses do not
matter, and the relative suppression of direct photoproduction is due to the
fact that, at LO, this is a pure CO process.

At $p_T=5$~GeV, single-resolved photoproduction is still overwhelming, as is
evident from Fig.~\ref{fig:xs}(b).
In Fig.~\ref{fig:xs}(b), the two pronounced maxima in the dotted line may be
traced to the $T$-matrix element of $\gamma q\to c\overline{c}[n]q$ that is
proportional to $e_c$, which contains a virtual gluon in the $t$ channel that
can become almost collinear with the incoming $q$ quark, and to the one that
is proportional to $e_q$, which contains a virtual $q$ quark in the $u$
channel that can become almost collinear with the incoming photon.

Furthermore, we observe from Fig.~\ref{fig:xs}(b) that the NLO correction to
direct photoproduction dramatically increases towards the forward and backward
directions.
This feature is also nicely exhibited in Fig.~\ref{fig:k}(b), where the NLO to
LO ratio is shown for $p_T=5$~GeV as a function of $y$ (solid line).
This is due to the finite remainders of the initial-state collinear
singularities that were absorbed into the photon PDFs.

In Fig.~\ref{fig:k}, two kinds of QCD correction ($K$) factor are studied (a)
for $y=0$ as a function of $p_T$ and (b) for $p_T=5$~GeV as a function of $y$.
The one that is defined as the ratio of the NLO and LO results of direct
photoproduction (solid lines) was already discussed above.
An alternative definition of $K$ factor is obtained by adding the LO result of
single-resolved photoproduction to both numerator and denominator, where it is
evaluated with the NLO and LO versions of $\alpha_s^{(n_f)}(\mu)$ and the
photon PDFs, respectively (dashed lines).
The rationale for this definition is the circumstance that the NLO result of
direct photoproduction and the LO result of single-resolved photoproduction
are interconnected by mass factorization as explained in
Section~\ref{sec:ana}.3.4.
For the reasons given in Section~\ref{sec:num}.1, this definition of $K$ factor
is slightly inconsistent and the resulting values should, therefore, be taken
with a grain of salt.
Since the LO result of single-resolved photoproduction dominates both
numerator and denominator of this $K$ factor, the latter takes moderate
values.
The fact that these values are even below unity in the considered regions of
phase space is due to the different choices of $\alpha_s^{(n_f)}(\mu)$ and the
photon PDFs for the single-resolved contribution in the numerator and
denominator of this $K$ factor.

\section{Conclusions}
\label{sec:con}

In the case of charmonium, the experimental verification of the NRQCD
factorization hypothesis in two-photon collisions at LEP2, photoproduction at
HERA, and hadroproduction at the Tevatron has now practically come to a halt,
one reason being that the available theoretical predictions for the production
of prompt $J/\psi$ mesons with finite values of $p_T$ in high-energetic
photon-photon, photon-hadron, and hadron-hadron collisions are only of LO and
thus suffer from considerable uncertainties, mostly from the dependences on 
the
renormalization and factorization scales and from the lack of information on
the nonperturbative MEs.

Having recently considered the cross sections of processes~(\ref{eq:ccg})
\cite{gg} and (\ref{eq:ccgg})--(\ref{eq:ccuu}) \cite{npb} in connection with
the associated production of prompt $J/\psi$ mesons with one or two identified
energetic hadron jets, respectively, in this paper, we took the next step by
studying at NLO the inclusive production of prompt $J/\psi$ mesons with finite
values of $p_T$.
This is the first time that an inclusive $2\to2$ process was treated at NLO in
the NRQCD factorization framework.
The technical difficulties that needed to be tackled include the treatment of
UV, collinear, soft, and Coulomb singularities, the NRQCD operator
renormalization, the mass factorization, and the analytic evaluation of
five-point one-loop integrals with UV, soft, and Coulomb singularities.

As for the real corrections, we employed the phase-space slicing method to
demarcate the regions of phase space containing soft and collinear
singularities from the hard regions, where the phase-space integrations were
carried out numerically.
We verified that the combined result is, to very good approximation,
independent of the choices of the cut-off parameters $\delta_i$ and
$\delta_f$, over an extended range of values.
We worked in dimensional regularization in connection with the
$\overline{\rm MS}$ renormalization and factorization schemes, so that our NLO
result depends on the QCD and NRQCD renormalization scales $\mu$ and
$\lambda$, respectively, and on the factorization scale $M$ connected with the
collinear splitting of the incoming photons into massless $q\overline{q}$
pairs.
While the $\mu$ and $\lambda$ dependences are formally cancelled up to terms
beyond NLO within direct photoproduction, the $M$ dependence is only
compensated by the LO cross section of single-resolved photoproduction.
By the same token, the strong $M$ dependence of the latter is considerably
reduced by the inclusion of our new result.
This is a crucial phenomenological merit of our work, as may be appreciated by
observing that, at LO, the overwhelming bulk of the cross section of prompt
$J/\psi$ production in two-photon collisions is due to single-resolved
photoproduction \cite{god,gg}.

The $K$ factor of direct photoproduction turned out to be very substantial at
large values of $p_T$ because fragmentation-prone channels start to open up at
NLO [see Fig.~\ref{fig:k}(a)].
In fact, at $p_T\gg2m$, the $p_T$ distribution of direct photoproduction at
NLO is rather similar to the LO result of single-resolved photoproduction both
in shape and normalization [see Fig.~\ref{fig:xs}(a)].

In order to complete the NLO treatment of prompt $J/\psi$ production in
two-photon collisions, we still need to evaluate the NLO corrections to
single- and double-resolved photoproduction.
Then, also prompt $J/\psi$ production in photoproduction at HERA and
hadroproduction at the Tevatron can be described at NLO.
This will provide a solid basis for an ultimate test of the NRQCD
factorization framework.

\bigskip
\noindent
{\bf Acknowledgements}
\smallskip

We thank S. Dittmaier for useful discussions about the treatment of the
five-point functions and F. Maltoni for a useful communication concerning the
identification of scales in Eq.~(\ref{eq:rge}).
This work was supported in part by the Deutsche Forschungsgemeinschaft through
Grant No.\ KN~365/1-1, by the Bundesministerium f\"ur Bildung und Forschung
through Grant No.\ 05~HT1GUA/4, and by Sun Microsystems through Academic
Equipment Grant No.~EDUD-7832-000332-GER.

\def\theequation{\Alph{section}.\arabic{equation}}
\begin{appendix}
\setcounter{equation}{0}

\section{Light-quark loop contribution}
\label{sec:lq}

The interference of the light-quark box amplitude ${\cal T}_q$ with the
tree-level one ${\cal T}_0$, which enters Eq.~(\ref{eq:vir}), reads
\begin{eqnarray}
2\re\left({\cal T}_0^*{\cal T}_q\right)&=&\frac{16e_q^2e_c^2e^4g_s^4
\left\langle{\cal O}[{}^3\!S_1^{(8)}]\right\rangle}{3m}
\left\{   \frac{stu
         [ s^2 + t^2 + 3tu + u^2 + 
           3s( t + u )  ] }{{\pi }^2
         {( s + t ) }^2{( s + u ) }^2
         {( t + u ) }^2}\right.\nonumber\\ 
  &&{}-    \frac{128m^4}{
         {( s + t ) }^3{( s + u ) }^3
         {( t + u ) }^3} 
         \left[ -4s^3tu( t + u )  - 
           4st^2u^2( t + u )   
    +       s^4( t^2 + u^2 )\right.\nonumber\\  
    &&{}+\left. t^2u^2( t^2 + u^2 )   
    +    s^2( t^4 - 4t^3u - 14t^2u^2 - 4tu^3 + 
              u^4 )  \right] 
         B_0(4m^2,0,0) \nonumber\\
 &&{}+  \frac{8s[ -4tu( t + u )  + 
           s( t^2 + u^2 )  ]}{( s + t ) 
         ( s + u ) {( t + u ) }^3}
         B_0(s,0,0) \nonumber\\
   &&{}+ \frac{8t[ s^2( t - 4u )  - 4su^2 + 
           tu^2 ] }{
         ( s + t ) {( s + u ) }^3
         ( t + u ) }B_0(t,0,0) \nonumber\\ 
  &&{}+    \frac{8u[ -4st^2 + t^2u + 
           s^2( -4t + u )  ] 
         }{{( s + t ) }^3
         ( s + u ) ( t + u ) }B_0(u,0,0) 
 \nonumber\\
 &&{}+ \frac{4s( 2s^2 + t^2 + u^2 ) 
         }{( s + t ) ( s + u ) 
     ( t + u ) }C_0(0,0,s,0,0,0)\nonumber\\  
    &&{}+  \frac{4t( s^2 + 2t^2 + u^2 ) 
        }{( s + t ) 
         ( s + u ) 
       ( t + u ) } C_0(0,t,0,0,0,0)\nonumber\\
 &&{}+ \frac{4u( s^2 + t^2 + 2u^2 ) 
         }{( s + t ) 
         ( s + u ) ( t + u ) }
      C_0(0,u,0,0,0,0)\nonumber\\ 
 &&{}- \frac{4( s^2 + 2t^2 + u^2 ) }{
         ( s + t ) ( t + u ) }
   C_0(4m^2,0,t,0,0,0)\nonumber\\
    &&{}-  \frac{4( s^2 + t^2 + 2u^2 ) }{
         ( s + u ) ( t + u ) }
       C_0(4m^2,0,u,0,0,0)\nonumber\\
 &&{}- \frac{4( 2s^2 + t^2 + u^2 ) }{
         ( s + t ) ( s + u ) }
       C_0(4m^2,s,0,0,0,0) \nonumber\\
 &&{}- \frac{4st( s^2 + t^2 ) 
         }{( s + t ) ( s + u ) 
         ( t + u ) }D_0(4m^2,0,0,0,t,s,0,0,0,0) 
  \nonumber\\
&&{}- \frac{4tu( t^2 + u^2 ) 
         }{( s + t ) ( s + u ) 
         ( t + u ) }D_0(4m^2,0,0,0,t,u,0,0,0,0)\nonumber\\
 &&{}-\left. 
      \frac{4su( s^2 + u^2 )}
{( s + t ) ( s + u ) ( t + u ) }D_0(4m^2,0,0,0,u,s,0,0,0,0)
 \right\},
\label{eq:lq}
\end{eqnarray}
where $B_0$, $C_0$, and $D_0$ denote the one-loop scalar two-, three-, and
four-point functions \cite{pas} in the notation of Ref.~\cite{Denner:kt}
(see also Ref.~\cite{bk}).
Since the $C_0$ and $D_0$ functions are IR divergent, we list them in
analytical form.
We have
\begin{eqnarray}
 B_0(s,0,0) &=& C_\epsilon\left(\frac{1}{\epsilon}-
\ln\frac{-\overline{s}}{m^2}+2\right),
\nonumber\\
 C_0(s,0,0,0,0,0) &=& C_\epsilon\frac{1}{s}\left[\frac{1}{\epsilon^2}
-\frac{1}{\epsilon}\ln\frac{-\overline{s}}{m^2}
+\frac{1}{2}\ln^2\frac{-\overline{s}}{m^2}-\frac{1}{2}\zeta(2)\right],
\nonumber\\
 C_0(4m^2,0,s,0,0,0) &=&-C_\epsilon\frac{1}{s-4m^2}
\left(\frac{1}{\epsilon}\ln\frac{\overline{s}}{4m^2}
+\frac{1}{2}\ln^2\frac{\overline{s}}{4m^2}
-\ln\frac{\overline{s}}{4m^2}\ln\frac{\overline{s}}{m^2}\right),
\nonumber\\
 D_0(4m^2,0,0,0,s,t,0,0,0,0) &=& C_\epsilon\frac{1}{st}\left[
\frac{2}{\epsilon^2}-\frac{2}{\epsilon}
\ln\frac{-\overline{s}\overline{t}}{4m^4}
+2\li\left(1-\frac{\overline{s}}{4m^2}\right)\right.
\nonumber\\
&&{}+\left.2\li\left(1-\frac{\overline{t}}{4m^2}\right)
+\ln^2\frac{-\overline{s}\overline{t}}{4m^4}-3\zeta (2)\right],
\end{eqnarray}
where $\zeta(2)=\pi^2/6$, $\bar{s}=s+i \epsilon$, $\bar{t}=t+i \epsilon$, and
$C_\epsilon$ is defined in Eq.~(\ref{eq::def_ceps}).
The result for $C_0(s,0,0,0,0,0)$ can be found, e.g., in
Ref.~\cite{Beenakker:2002nc}, and those for $C_0(4m^2,0,s,0,0,0)$ and
$D_0(4m^2,0,0,0,s,t,0,0,0,0)$ can be extracted from Ref.~\cite{Bern:1993kr}.

\end{appendix}

\newpage
 
\newpage

\newpage

\newpage

\newpage

\newpage

\newpage
\begin{figure}[ht]
\begin{center}
\epsfig{figure=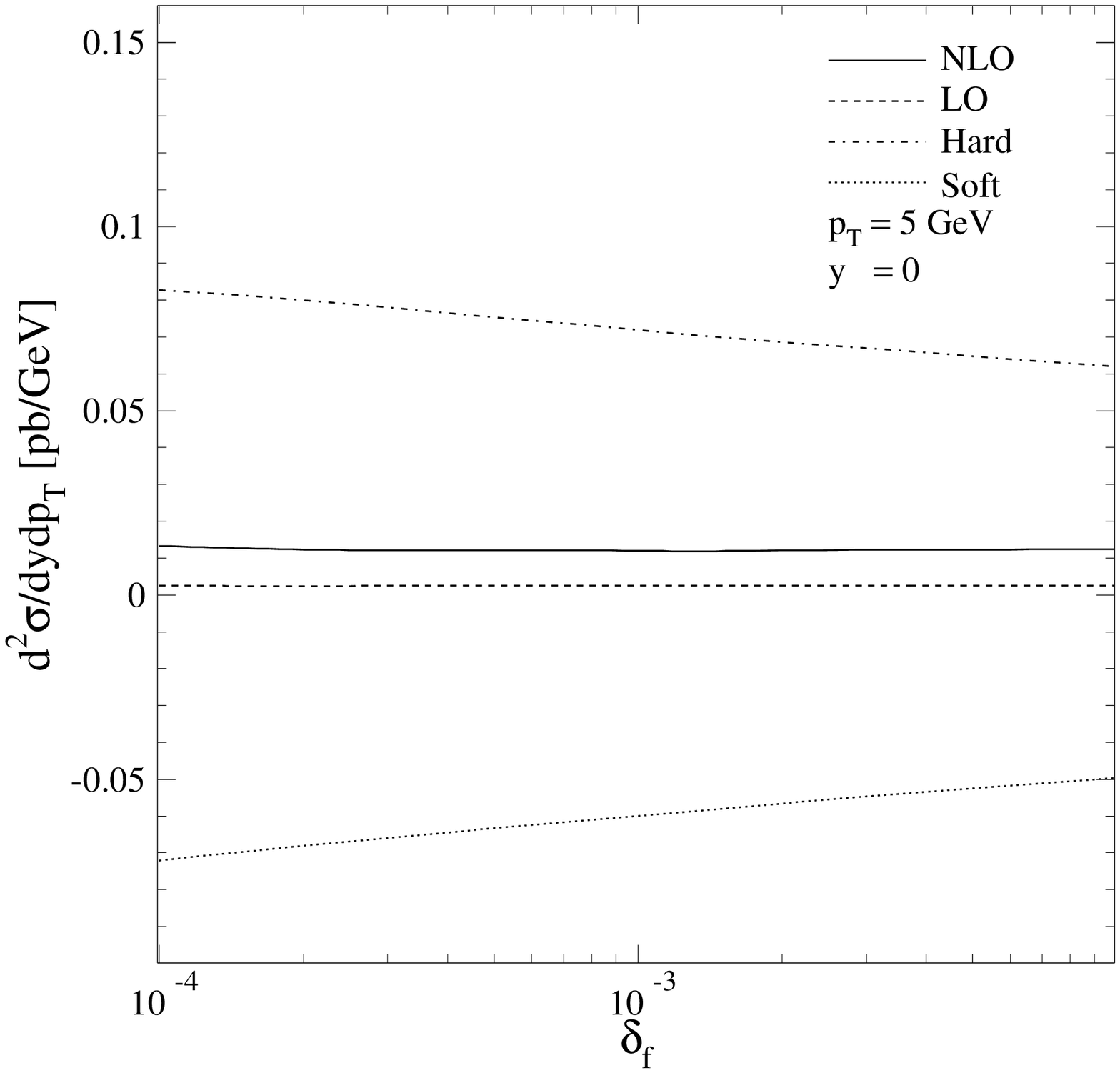,width=\textwidth}
(b)\\
Fig.~\ref{fig:cut} (continued).
\end{center}
\end{figure}

\newpage

\newpage

\newpage

\newpage

\newpage
\begin{figure}[ht]
\begin{center}
\epsfig{figure=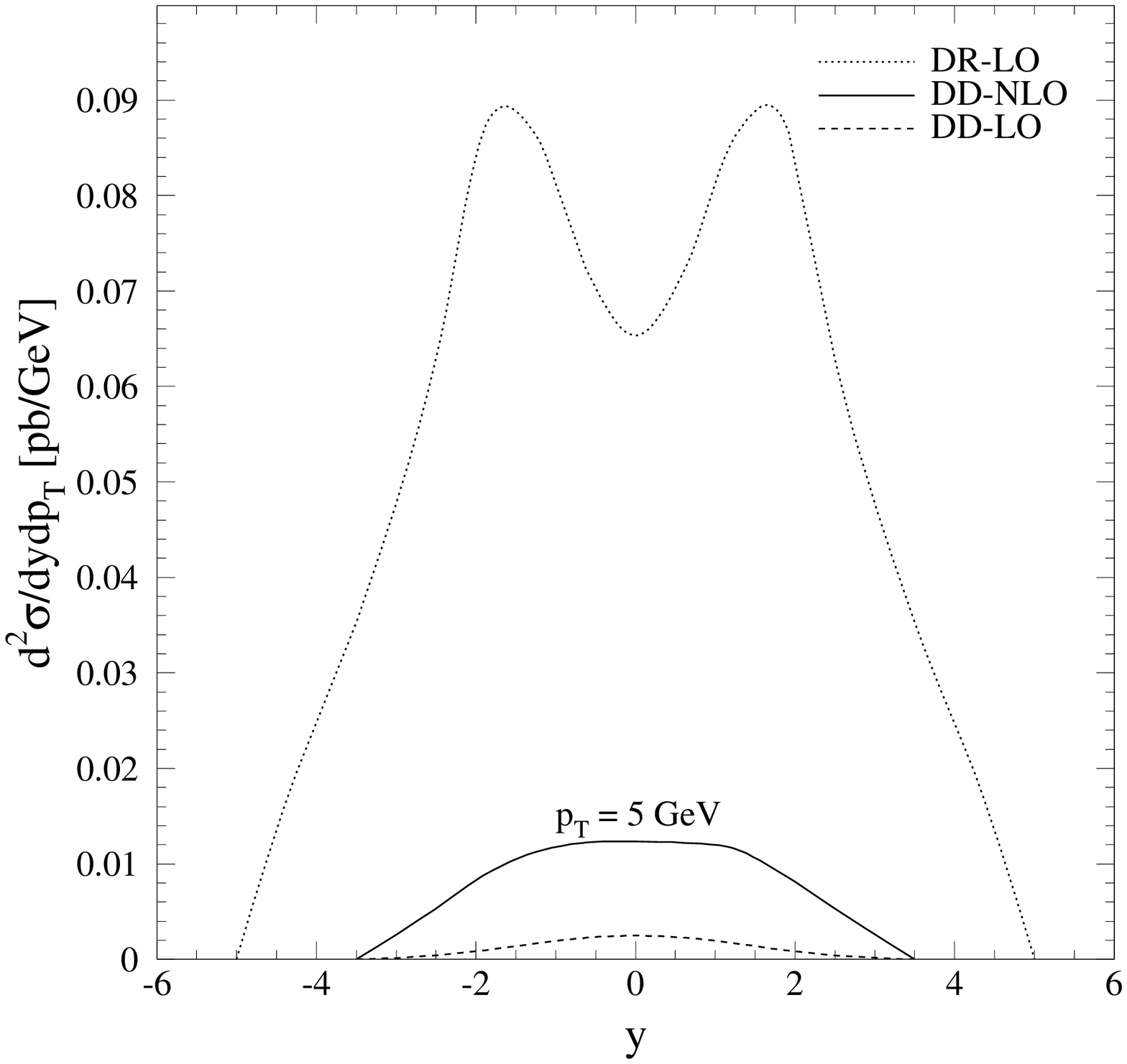,width=\textwidth}
(b)\\
Fig.~\ref{fig:xs} (continued).
\end{center}
\end{figure}

\newpage

\newpage
\begin{figure}[ht]
\begin{center}
\epsfig{figure=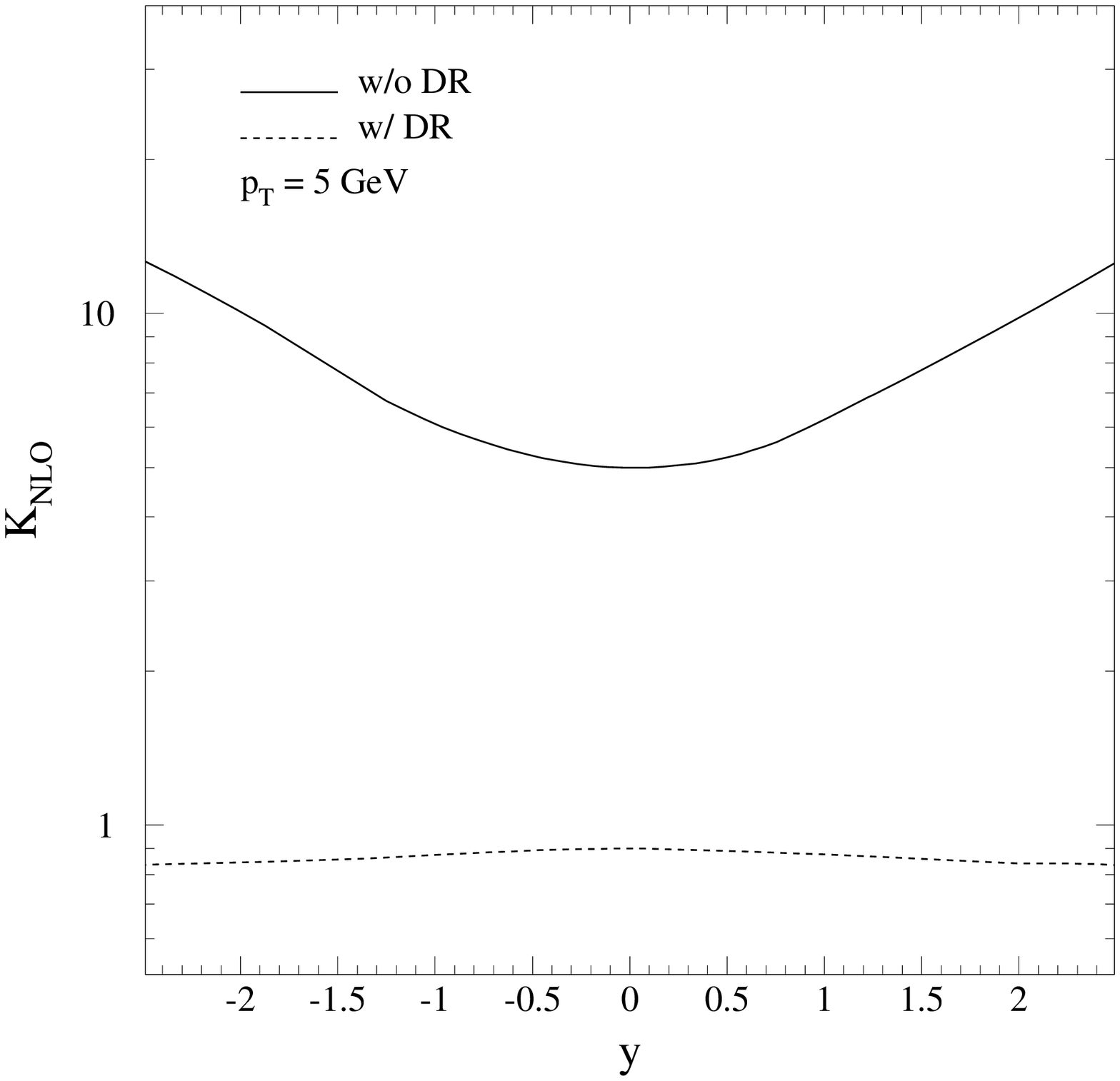,width=\textwidth}
(b)\\
Fig.~\ref{fig:k} (continued).
\end{center}
\end{figure}

\end{document}